\documentclass[aps,prc,twocolumn,times,graphicx,tighten,showpacs]{revtex4}
\usepackage[dvips]{graphicx}
\usepackage[dvips]{color}
\usepackage{amsmath}

\newcommand{\bea}{\begin{eqnarray}}
\newcommand{\eea}{\end{eqnarray}}
\newcommand{\be}{\begin{equation}}
\newcommand{\ee}{\end{equation}}
\newcommand{\np}{{\bf p}}

\newcommand{\nh}{{\bf h}}
\newcommand{\nk}{{\bf k}}
\newcommand{\nl}{{\bf l}}
\newcommand{\nx}{{\bf x}}
\newcommand{\nq}{{\bf q}}
\newcommand{\ntau}{{\bf \tau}}
\newcommand{\Qbar}{\not{\!Q}}

\newcommand{\kbar}{\not{\!k}}

\newcommand{\Pbar}{\not{\!P}}
\newcommand{\nsigma}{\mbox{\boldmath $\sigma$}}
\newcommand{\tauvec}{\mbox{\boldmath $\tau$}}
\newcommand{\Ivec}{\mbox{\boldmath $I$}}

\newcommand{\sumint}{\sum\kern -3.5ex \int\kern 1.0ex}

\newlength\dlf  

\begin{document}

\title{Relativistic model of 2p-2h meson exchange currents in (anti)neutrino
scattering
}

\author{
I. Ruiz Simo$^a$,
J.E. Amaro$^a$,
M.B. Barbaro$^{b,c}$,
A. De Pace$^b$,
J.A. Caballero$^d$,
T.W. Donnelly$^e$,
}

\affiliation{$^a$Departamento de F\'{\i}sica At\'omica, Molecular y Nuclear,
and Instituto de F\'{\i}sica Te\'orica y Computacional Carlos I,
Universidad de Granada, Granada 18071, Spain}

\affiliation{$^b$INFN, Sezione di Torino, Via P. Giuria 1, 10125 Torino, Italy}
  
\affiliation{$^c$Dipartimento di Fisica, Universit\`a di Torino 
 Via P. Giuria 1, 10125 Torino, Italy}

\affiliation{$^d$Departamento de F\'{\i}sica At\'omica, Molecular y Nuclear,
Universidad de Sevilla, Apdo.1065, 41080 Sevilla, Spain}

\affiliation{$^e$Center for Theoretical Physics, Laboratory for Nuclear
  Science and Department of Physics, Massachusetts Institute of Technology,
  Cambridge, MA 02139, USA}

\date{\today}


\begin{abstract}

We develop a model of relativistic, charged meson-exchange currents
(MEC) for neutrino-nucleus interactions.  The two-body current is the
sum of seagull, pion-in-flight, pion-pole and $\Delta$-pole
operators. These operators are obtained from the weak pion-production
amplitudes for the nucleon derived in the non-linear $\sigma$-model
together with weak excitation of the $\Delta(1232)$ resonance and its
subsequent decay into $N\pi$.  With these currents we compute the five
2p-2h response functions contributing to $(\nu_l,l^-)$ and
$(\overline{\nu}_l,l^+)$ reactions in the relativistic Fermi gas
model. The total current is the sum of vector and axial two-body
currents. The vector current is related to the electromagnetic MEC
operator that contributes to electron scattering. This allows one to check our
model by comparison with the results of De Pace {\em et al.,} Nuclear
Physics A 726 (2003) 303.  Thus our model is a natural extension of
that model to the weak sector with the addition of the axial
MEC operator.  The dependences of the response functions on several
ingredients of the approach are analyzed. Specifically we discuss
relativistic effects, quantify the size of the direct-exchange
interferences, and the relative importance of the axial versus vector
current. 

\end{abstract}


\maketitle

\section{Introduction}

Modern accelerator-based neutrino experiments use nuclear targets to
extract neutrino oscillation parameters. The charged current
quasielastic (CCQE) process $(\nu_l,l)$ dominates at the typical
energies of 1 GeV of the incident neutrino flux, whose reconstruction
from the final state requires some assumptions about the initial
and final nuclear states. Within the crude approximation that
the neutrino interacts with a
(bound) nucleon at rest, the neutrino energy can be calculated using
only the outgoing lepton kinematics.  A recent review of the present
understanding of the neutrino-nucleus interaction and its effects on
the neutrino energy reconstruction is presented in \cite{Mos16}.
Besides, the uncertainties in the neutrino-nucleon interaction, 
for instance the present limited knowledge of the axial form factor~\cite{Bha15,Ama16}, 
the nuclear many-body effects and the final-state
interactions which are not experimentally distinguishable
complicate the energy reconstruction.

The events in which two or more nucleons are ejected have
been suggested to be important, based on several calculations of the QE neutrino cross
section~\cite{Mar09,Mar10,Ama11,Nie11,Nie12,Ama12,Gra13}.
The contribution of two-particle-two-hole (2p-2h) excitations are believed to be
essential for a proper description of recent neutrino 
experiments~\cite{Agu10,Nak11,And12,Abe13,Fio13,Abe14,Wal15,Ank16,Rod16}.  At the
intermediate momentum transfers typical of these experiments,
relativistic effects have to be taken into account, not only in the
kinematics and nuclear wave functions,
but also in the current operators.

Several calculations of 2p-2h excitations in neutrino scattering have
been reported, each of them relying on different assumptions and
approximations.  In particular, the model in \cite{Mar09,Mar10} is based
on a non-relativistic treatment of meson-exchange currents (MEC) and
correlations, with some relativistic corrections added, and the axial
MEC contribution estimated from the well-known vector operator.  The
model in \cite{Nie11}, on the other hand, is relativistic and uses some
approximations to compute the momentum-space integrals of the 2p-2h
matrix elements; it also neglects some contributions, 
in particular the direct/exchange interference.
Both of the
  above calculations use the Fermi gas model to compute the 2p-2h
  matrix elements.  Worth mentioning are also some recent efforts
towards alternative approaches to the problem, where a factorization
ansatz is assumed in order to account 
for one- and two-body current contributions~\cite{Ben15,Roc15}.

Within the super-scaling approach (SuSA) the MEC contribution to the neutrino
cross section was estimated~\cite{Ama11} from the electromagnetic (em)
2p-2h transverse response model of~\cite{DePace:2003spn,DePace04}, by
neglecting the axial component. This model is fully relativistic,
includes all the interference diagrams and evaluates the
seven-dimensional integrals without approximations. 

Our goal in this paper is to extend the model of~\cite{DePace:2003spn} 
by including the axial current. This study will be applied to the analysis of 
neutrino-nucleus scattering processes in a forthcoming publication~\cite{Guille2} based on 
our recent investigation of electron scattering reactions making use of the superscaling approach~\cite{Meg15,Meg16} 
(see also \cite{SuSAv2} for details).

The relativistic Fermi gas (RFG) is the simplest model that
allows a complete and fully relativistic calculation of 2p-2h effects.
This requires one to compute the spin-isospin traces of all of the
many-body diagrams. In electron scattering they involve more than
100,000 terms that were evaluated in \cite{DePace:2003spn} with
subsequent 7D integrals. To extend this procedure to neutrino
scattering implies adding the new axial MEC operator, and computing five
nuclear response functions. In addition to the pure vector pieces $(VV)$
in the squared amplitudes, new axial $(AA)$ and interference $(VA)$
contributions appear, thus increasing the number of traces to
compute. The resulting number of terms would make this procedure
almost intractable.  Therefore, instead of computing analytically the
traces with Dirac matrices algebra, here we follow the 
approach of \cite{Amaro:2010iu} by computing numerically the spin
traces.

We first introduce briefly in Sect. II the formalism of neutrino
scattering.  In Sect. III we provide the expressions for the
relativistic MEC matrix elements in the 2p-2h channel. The novel piece
is the inclusion of the relativistic axial MEC, deduced from the weak
pion production amplitudes of \cite{Hernandez:2007qq}. 

The expressions of the 2p-2h response
functions in terms of the current matrix elements for the separate
isospin channels are given in Sect. IV. 

An important point that deserves clarification concerns the ambiguities
between MEC and Delta peak contributions.  The $\Delta$ peak is the main
contribution to the pion production cross section.  Inside the nucleus
the $\Delta$ can decay into one nucleon that re-scatters producing
two-nucleon emission, therefore this channel should be considered part
of the 2p-2h channel.  In Sect. V we describe how we treat this
issue within our approach.  

In Sect. VI we present and discuss the results for the 2p-2h weak and
electromagnetic nuclear response functions for several kinematics and
we analyze the importance of the different contributions and
ingredients of the model.  Finally in Sect. VII we summarize our
findings and draw our conclusions.

\section{Formalism of neutrino scattering}

This formalism can be applied to both neutrino $(\nu_l,l^-)$ and
antineutrino $(\overline{\nu}_l,l^+)$ CC reactions in nuclei. The
incident and scattered leptons have momenta $K^\mu=(\epsilon,\nk)$ and
$K'{}^\mu=(\epsilon',\nk')$, respectively.  The
four-momentum transfer is $Q^{\mu}=(\omega,\nq)=(K-K')^{\mu}$, where
$\omega$ is the energy transfer and $\nq$ is the momentum transfer.
We choose the $z$
axis along the $\bf q$ direction. 
The double-differential cross section is~\cite{Ama05}
\begin{equation}
\frac{d\sigma}{d\Omega'd\epsilon'}
= \sigma_0 {\cal S_{\pm}},
\label{sigma}
\end{equation}
where $\sigma_0$ is a kinematical factor including the weak
  couplings defined in \cite{Ama05}.  The nuclear structure function
${\cal S_{\pm}}$ is the linear combination of five response functions
\begin{eqnarray}
{\cal S_{\pm}}
&=& 
\widetilde{V}_{CC} R^{CC}
+ 2 \widetilde{V}_{CL} R^{CL}
+ \widetilde{V}_{LL} R^{LL}
\nonumber\\
&&
+ \widetilde{V}_{T} R^{T}
\pm 2 \widetilde{V}_{T'} R^{T'},
\label{Spm}
\end{eqnarray}
where the sign of the last term is positive for neutrinos and negative
for antineutrinos.
The $\widetilde{V}_K$'s are kinematical factors defined in \cite{Ama05}.
In this paper we are interested in the five nuclear response functions
\begin{eqnarray}
R^{CC} &=& W^{00} \label{rcc} \\
R^{CL} &=& -\frac12\left(W^{03}+ W^{30}\right) \\
R^{LL} &=& W^{33}  \\
R^{T} &=& W^{11}+ W^{22} \\
R^{T'} &=& -\frac{i}{2}\left(W^{12}- W^{21}\right). \label{rtp}
\end{eqnarray}
The hadronic tensor $W^{\mu\nu}$ 
is calculated in a relativistic Fermi gas (RFG) model
with Fermi momentum $k_F$. 
The final states can be excitations of the $n$p-$n$h kind. 
Thus the hadronic tensor can be expanded as the sum of 
one-particle one-hole (1p-1h) , two-particle two-hole (2p-2h), 
plus additional channels 
\begin{equation}
W^{\mu\nu} = W^{\mu\nu}_{1p1h} + W^{\mu\nu}_{2p2h} + \cdots
\end{equation}
In the impulse approximation the 1p-1h channel
gives the well-known response functions of the RFG~\cite{Ama05}. 
Here we focus on
the 2p-2h channel, with two nucleons 
with momenta $\np'_1$ and $\np'_2$
above the Fermi momentum, $p'_i>k_F$,
and two hole states with momenta $\nh_1$ and $\nh_2$ 
below the Fermi momentum, $h_i<k_F$.
The spin (isospin) indices are $s'_i$ ($t'_i$) and
$s_i$ ($t_i$), respectively.

The  2p-2h
 hadronic tensor in the RFG model
 is proportional to the volume $V$ of the system, 
which for symmetric nuclear matter, $Z=N=A/2$, is  
 $V=3\pi^2 Z/k_F^3$.  It is given by
\begin{eqnarray}
W^{\mu\nu}_{2p-2h}
&=&
\frac{V}{(2\pi)^9}\int
d^3p'_1
d^3h_1
d^3h_2
\frac{M^4}{E_1E_2E'_1E'_2}
\nonumber \\ 
&&
r^{\mu\nu}(\np'_1,\np'_2,\nh_1,\nh_2)
\delta(E'_1+E'_2-E_1-E_2-\omega)
\nonumber\\
&&
\Theta(p'_1,p'_2,h_1,h_2),
\label{hadronic}
\end{eqnarray}
where 
$\bf p'_2= h_1+h_2+q-p'_1$
is fixed by  momentum conservation,
$M$ is the nucleon mass, the energies $E_i$ and $E'_i$ are the on-shell energies of the holes
and particles,
and 
\begin{eqnarray}
\Theta(p'_1,p'_2,h_1,h_2)
&\equiv&
\theta(p'_2-k_F)
\theta(p'_1-k_F)
\nonumber\\
&&\times 
\theta(k_F-h_1)
\theta(k_F-h_2).
\end{eqnarray}
The non-trivial part of the calculation is contained in the 
function $r^{\mu\nu}(\np'_1,\np'_2,\nh_1,\nh_2)$, which represents
the elementary hadronic tensor for the basic 2p-2h transition,
with the given initial and final momenta, summed over spin and
isospin projections
\begin{eqnarray}
\lefteqn{r^{\mu\nu}(\np'_1,\np'_2,\nh_1,\nh_2)=}
 \nonumber\\ 
&& \frac{1}{4}\sum_{s_1s_2s'_1s'_2}
\sum_{t_1t_2t'_1t'_2}
j^{\mu}(1',2',1,2)^*_A
j^{\nu}(1',2',1,2)_A.
 \nonumber\\ 
\label{elementary}
\end{eqnarray}
This elementary hadronic tensor is written in terms of the two-body
MEC antisymmetrized matrix element
 $j^{\mu}(1',2',1,2)_A$ (we use the definition given in
Eq. (17) of \cite{Ama02}).  The factor $1/4$ accounts for the
antisymmetry of the 2p-2h wave function to avoid double counting. 

The above sum over isospin combines all the possible charge channels
in the final state, corresponding to emission of $pp$, $nn$ and $pn$
pairs. In our formalism, discussed below, we separate the contributions
of these charge states to the response functions. Although we present
results for the total contribution, having the possibility to separate
the isospin contributions will allow us to apply the formalism to
asymmetric nuclei $N\ne Z$. This will be of interest~\cite{Ada15} for
neutrino experiments based, for instance, on $^{40}$Ar, $^{56}$Fe or $^{208}$Pb.  

To compute the hadronic tensor in Eq.~(\ref{hadronic})
 we took advantage of the symmetry imposed by the choice of having the $z$
axis along the ${\bf q}$ direction. 
 Then the rotational symmetry of the
response functions, Eqs.~(\ref{rcc}--\ref{rtp}), 
 around the $\nq$ direction allows us to integrate
over one of the azimuthal angles. We
choose $\phi'_1=0$ and multiply the responses by a factor $2\pi$.
Furthermore, the energy delta function enables analytical integration
over $p'_1$. The 2p-2h integral is then reduced to 7 dimensions.

In this paper we evaluate the resulting 7D integral using the
numerical method described in \cite{Ruiz14,Ruiz14b}.  It is
useful to calculate the hadronic tensor expected for $r^{\mu\nu}=1$
({\it i.e.} that arising from phase-space alone).  This was done in
\cite{Ruiz14} in the laboratory frame, and in \cite{Ruiz14b}
in the hadronic center-of-mass (CM) system.  These results will be
modified here when including the effects of the two-body physical
current.  
A related analysis was done in \cite{Lal12,Lal12b} with
a pure 2p-2h phase space alone, fitted to the experimental cross
section, where the effects of the physics in the tensor $r^{\mu\nu}$
were not taken into account.

\section{Electroweak meson-exchange currents}

The evidence for a pion-exchange contribution to the axial current of
nuclei is well-known from weak processes such as $\beta$ decay,
$\mu$-capture or solar proton burning $pp\rightarrow de^+\nu_e$, generating
several theoretical studies \cite{Eri88,Ris89,Bar16}, all of which are focused 
on low-energy processes where a non-relativistic description is
adequate, and where one starts from a non-relativistic
current operator. However, for the energies involved in modern
neutrino experiments a relativistic approach is mandatory, and accordingly we
start with a fully relativistic operator.

\begin{figure}[tph]
\begin{center}
\includegraphics[scale=0.77,bb=160 200 454 686]{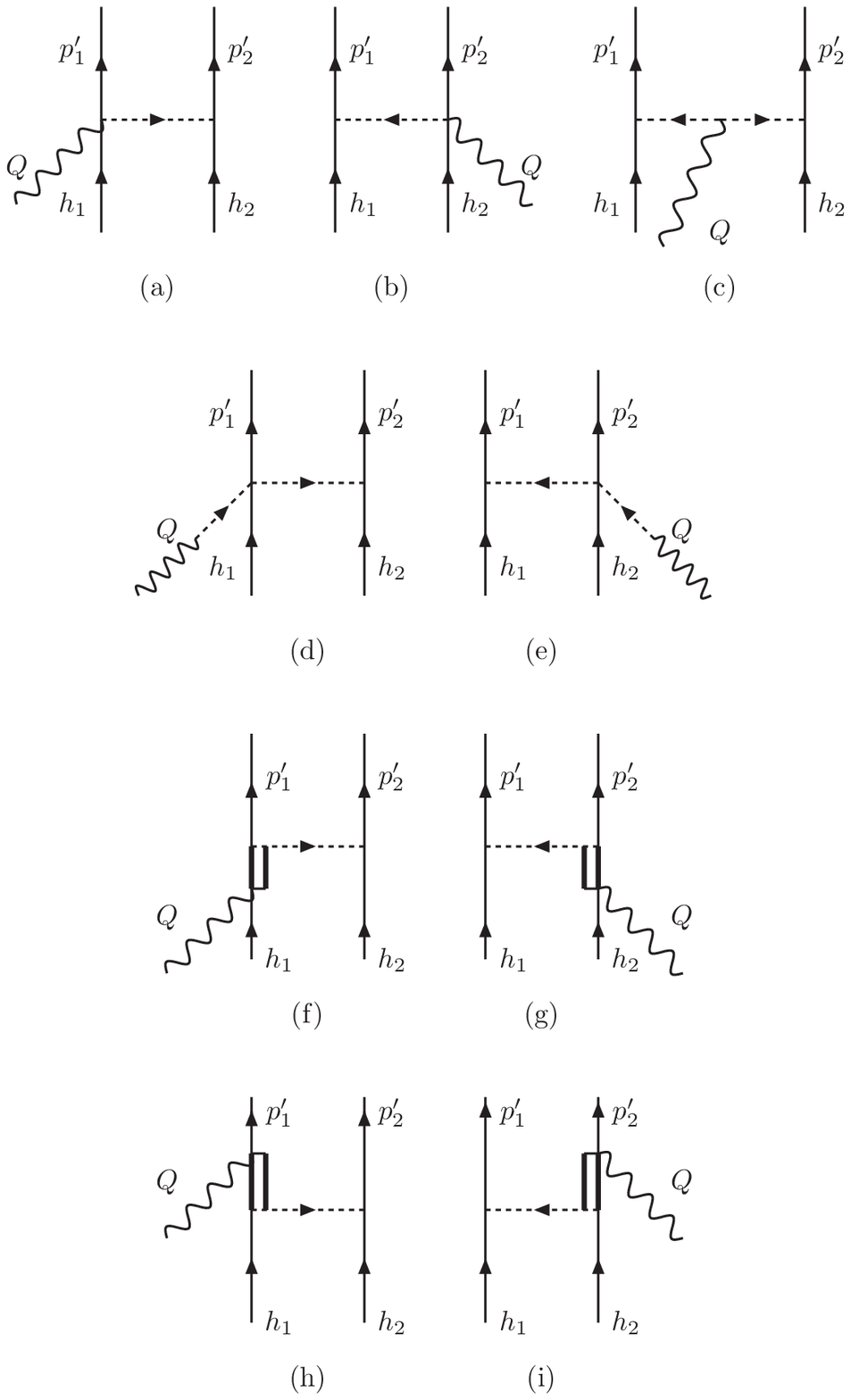}
\caption{ Feynman diagrams of the MEC considered in the present study,
  including the seagull (a,b), pion-in-flight (c), pion-pole (d,e),
  and $\Delta$ pole (f--i).  
}
\label{diagrams}
\end{center}
\end{figure}

In this section we describe our model for relativistic weak meson-exchange 
currents.  We work at tree level including only one-pion
exchange. The relativistic electromagnetic MEC has been widely studied
previously for intermediate-energy electron scattering and 
expressions have been given for instance in~\cite{Dek94,DePace:2003spn,Amaro:2010iu}. 
To obtain the weak two-body
current we need the relativistic axial contribution.  To our knowledge
this current has not been written explicitly in the literature, so
this is one of the novelties of this work. We start from the weak pion
production model of \cite{Hernandez:2007qq}, based on the
non-linear $\sigma$-model.  We take the pion-production amplitudes
from the nucleon given there, and we couple a second nucleon-line to
the emitted pion.  The resulting MEC operator is written as the sum of
four contributions, denoted as seagull, pion-in-flight, pion-pole and
Delta-pole
\begin{eqnarray}
 j^\mu_{\rm MEC}=
 j^\mu_{\rm sea}+
 j^\mu_{\rm \pi}+
 j^\mu_{\rm pole}+
 j^\mu_{\rm \Delta} .
\end{eqnarray}
The corresponding Feynman diagrams are given in Fig.~\ref{diagrams}.

In this work we do not include the so-called nucleon-pole
contributions. These can be considered a part of the nucleon
correlations that contributes to the nuclear spectral function and final-state 
interactions (FSI), and are not considered genuine
meson-exchange currents. Besides, the corresponding diagrams produce
divergences in the quasielastic region and some kind of 
regularization~\cite{Alb84,Amaro:2010iu,Alb91} or subtraction of self-energy 
diagrams~\cite{Gil97,Nie11} are required.  The effect of correlations is taken
into account, at least partially, in the one-body cross section by
using a spectral function for the nucleon in the medium~\cite{Nie04,Ben08} 
or alternatively, with the super-scaling approach~\cite{Amaro:2004bs,Meg15}.

Each one of the four MEC operators can be decomposed as a sum of
vector $(V)$ and axial-vector $(A)$ currents. The vector operators
also contribute to electron scattering and are constrained by
electromagnetic probes, while the axial ones only appear in weak
processes like neutrino scattering.

\subsection{Seagull current}
The weak seagull current depicted in Fig.~\ref{diagrams}, diagrams (a)
and (b), can be written
as:
 \begin{equation}
 j^\mu_{\rm sea}= \left(I_V\right)_{\pm}
J^\mu_{\rm sea},
\label{seacur}
\end{equation}
where 
\begin{equation}
\left(I_V\right)_{\pm}= (I_V)_x\pm i (I_V)_y
\end{equation}
stands for the $\pm$-component of the
two-body isovector operator
\begin{equation}
\Ivec_V = 
i \left[\tauvec(1) \times\tauvec(2)\right]
\label{isospin_seagull}
\end{equation}
and $J^\mu_{\rm sea}$ is the isospin-independent seagull current,
given as the sum of $V$ and $A$ components  
\begin{equation}
J^\mu_{\rm sea}
=
\left(J^\mu_{\rm sea}\right)_V  + \left(J^\mu_{\rm sea}\right)_A.
\end{equation}
They have  been derived from
the contact term (CT) of the pion neutrino-production 
amplitudes from \cite{Hernandez:2007qq}, and can be written as
\begin{eqnarray}
 \left(J^\mu_{\rm sea}\right)_V&=&\frac{f^2_{\pi NN}}{m^2_\pi}
 F^V_1(Q^2) V^{(s^\prime_1,s_1)}_{\pi NN}(\np^\prime_1,\nh_1)
 \nonumber\\
&\times& \bar{u}_{s^\prime_2}(\np^\prime_2)
\gamma_5 \gamma^\mu u_{s_2}(\nh_2)
-(1\leftrightarrow2)\label{electro_seagull}
\\
 \left(J^\mu_{sea}\right)_A&=&\frac{f^2_{\pi NN}}{m^2_\pi} \frac{1}{g_A}
 V^{(s^\prime_1,s_1)}_{\pi NN}(\np^\prime_1,\nh_1)
 F_\rho\left(k_{22}^2\right)\nonumber\\
 &\times& \bar{u}_{s^\prime_2}(\np^\prime_2)\,\gamma^\mu\, 
 u_{s_2}(\nh_2) -  (1\leftrightarrow2).\label{axial_seagull}
\end{eqnarray}
In these equations:
\begin{itemize}

\item
The coupling constant $f_{\pi NN}/m_\pi$
comes from the $W^{\pm}\pi$NN and $\pi$NN vertices. 
In the $A$ component the Goldberger-Treiman relation has been applied to
the amplitudes of \cite{Hernandez:2007qq}
\begin{equation}\label{Goldberger_Treiman}
\frac{g_A}{2f_\pi}=\frac{f_{\pi NN}}{m_\pi},
\end{equation}
with $g_A=1.26$ and $f_\pi=93$ MeV being, respectively, 
the nucleon axial coupling and the pion decay constant.

\item
The same form factors as in \cite{Hernandez:2007qq} are used. $F^V_1(Q^2)$
is the vector nucleon form factor, and  $F_\rho(k^2)$ 
accounts for the $\rho$-meson dominance
of the $\pi\pi$NN coupling.

\item
The pion four-momenta $k_{ij}$ are the differences
between the final and initial nucleon four-momenta in the $\pi$NN vertex, {\it i.e.},
\begin{equation}
k_{ij}=p^{\prime}_i-h_j, \quad i,j=1,2.
\end{equation}

\item
The function
\begin{equation}
V^{(s^\prime_i,s_j)}_{\pi NN}(\np^\prime_i,\nh_j)=
\frac{\bar{u}_{s^\prime_i}(\np^\prime_i)\,\gamma_5\kbar_{ij} \, u_{s_j}(\nh_j)}{k^2_{ij}-m^2_\pi}
\end{equation}accounts for the propagation and subsequent absorption of the 
exchanged pion and includes
the pion propagator and the $\pi NN$ vertex
(see Fig.~\ref{diagrams}, diagrams (a),(b)).
Note that for on-shell nucleons the following relation can be used to simplify the expression of $V_{\pi NN}$
\begin{eqnarray}
&&\bar{u}_{s^\prime_i}(\np^\prime_i) \gamma_5\kbar_{ij} \, u_{s_j}(\nh_j)
= -2m \bar{u}_{s^\prime_i}(\np^\prime_i) \gamma_5  u_{s_j}(\nh_j).
\end{eqnarray}
\item
The Cabibbo angle $\theta_c$, which was present in
the amplitudes of \cite{Hernandez:2007qq}, has been
factorized out from the weak currents, and has been 
included in the definition of the factor $\sigma_0$ 
in Eq.~(\ref{sigma}).

\item
The shorthand notation $(1\leftrightarrow2)$ means to
interchange the ordering in the labels of the two nucleons' spins and
momenta, but not the isospins. Note that the $j^{\mu}_{\rm sea}$ current has to be
understood as an operator in isospin space and a matrix element in
spin-coordinate space.  The symmetrization of the above operator 
is automatically taken into account
 due to the antisymmetry of the isovector
isospin operator $\left(I_V\right)_{\pm}$ under the exchange of 
$(1\leftrightarrow 2).$ 

\end{itemize}

Finally, note that the electromagnetic current operator can be obtained
from the above equations by keeping only the $V$ current and taking the
$z$ component of the isospin operator
\begin{equation}
(I_V)_\pm \rightarrow (I_V)_z = i\left[\tauvec(1) \times\tauvec(2)\right]_z .
\end{equation}
The resulting electromagnetic seagull current is in agreement with
previous expressions~\cite{VanOrden:1980tg,DePace:2003spn,Amaro:2010iu}.

 \subsection{Pion-in-flight term}

 The weak pion-in-flight current, depicted in diagram (c) of
 Fig.~\ref{diagrams}, can be expressed in a similar way to the seagull
 operator, but with a vanishing axial part:
  \begin{eqnarray}
 j^\mu_{\pi}&=& \left(I_V\right)_\pm\,J^\mu_{\pi}
 \label{pionic_current1} \\
J^\mu_{\pi}  &=& \left(J^\mu_{\pi}\right)_V + \left(J^\mu_{\pi}\right)_A
\\
 \left(J^\mu_{\pi}\right)_V&=&\frac{f^2_{\pi NN}}{m^2_\pi}
 F^V_1(Q^2) 
V^{(s^\prime_1,s_1)}_{\pi NN}(\np^\prime_1,\nh_1)
\nonumber\\
 &\times&
 V^{(s^\prime_2,s_2)}_{\pi NN}(\np^\prime_2,\nh_2)
\left(k^\mu_{11}-k^\mu_{22}\right)\label{pionic_current2}
\\
 \left(J^\mu_{\pi}\right)_A&=& 0.
 \end{eqnarray}

Equation (\ref{pionic_current2}) reproduces the well-known expression for the
pion-in-flight electromagnetic MEC taking the $z$ component of the
isospin operator~\cite{VanOrden:1980tg,DePace:2003spn,Amaro:2010iu}.
It corresponds to the so-called PF piece of the pion production amplitudes of \cite{Hernandez:2007qq}.

\subsection{Pion-pole term}

At variance with the pion-in-flight current, the pion-pole term 
(diagrams (d) and (e) of Fig.~\ref{diagrams}) has
only the axial component and therefore it is absent in the
electromagnetic case. This new contribution could be considered as the
``axial counterpart'' of the pion-in-flight term, in the sense that it
contains two pion propagators and is proportional to
$k_{11}+k_{22}=Q$.  
The expression for this current is:
\begin{eqnarray}
j^\mu_{\rm pole}&=&\left(I_V\right)_{\pm} J^\mu_{\rm pole}
\label{polecur}
\\
J^\mu_{\rm pole}&=&
\left(J^\mu_{\rm pole}\right)_V + \left(J^\mu_{\rm pole}\right)_A 
\\
\left(J^\mu_{\rm pole}\right)_V &=& 0
\\ 
\left(J^\mu_{\rm pole}\right)_A 
&=&\frac{f^2_{\pi NN}}{m^2_\pi}
\frac{1}{g_A}F_\rho\left(k_{11}^2\right)
V^{(s^\prime_2,s_2)}_{\pi NN}(\np^\prime_2,\nh_2)\nonumber\\
&\times&Q^\mu\;
\frac{\bar{u}_{s^\prime_1}(\np^\prime_1)\Qbar\;
 u_{s_1}(\nh_1)}{Q^2-m^2_\pi}-(1\leftrightarrow2) .
\end{eqnarray}
Note the similarity with the axial part of the seagull current because
it has the same form factor and it contains a factor $1/g_A$.
Since it is proportional to $Q^\mu$, this current only contributes to
the longitudinal and time components of the hadronic tensor.

\subsection{$\Delta(1232)$ term}

The $\Delta$-pole terms correspond
in Fig.~\ref{diagrams} to diagrams (f,g) for the
forward and (h,i) for the backward $\Delta$ propagations, respectively.
We start from the $\Delta$-pole and the crossed-$\Delta$-pole
pion-production  amplitudes of \cite{Hernandez:2007qq}.
Attaching a second nucleon which absorbs the pion, we obtain the following 
currents
\begin{align}
j^\mu_{\Delta}&=j^\mu_{\Delta,\rm forw}+
j^\mu_{\Delta,\rm back}
\\
j^\mu_{\Delta,\rm forw}&=-\frac{f^* f_{\pi NN}}{m^2_\pi}\,
\sqrt{3}
\left(U^{\rm forw}\right)_{t'_1 t'_2; t_1 t_2}
\; 
V^{(s^\prime_2,s_2)}_{\pi NN}(\np^\prime_2,\nh_2)
\nonumber\\
&\times k^\alpha_{22}\,\bar{u}_{s^\prime_1}(\np^\prime_1)G_{\alpha\beta}(h_1+Q)
\Gamma^{\beta\mu}(h_1,Q)u_{s_1}(\nh_1)
\nonumber\\
&+(1\leftrightarrow2)
\label{delta_forward}
\\
j^\mu_{\Delta,\rm back}&=-\frac{f^* f_{\pi NN}}{m^2_\pi}\,
\sqrt{3}\left(U^{\rm back}\right)_{t'_1 t'_2; t_1 t_2}\; 
V^{(s^\prime_2,s_2)}_{\pi NN}(\np^\prime_2,\nh_2)
\nonumber\\
&\times k^\beta_{22}\,\bar{u}_{s^\prime_1}(\np^\prime_1)
\hat{\Gamma}^{\mu\alpha}(p^\prime_1,Q)
G_{\alpha\beta}(p^\prime_1-Q)u_{s_1}(\nh_1)
\nonumber\\
&+(1\leftrightarrow2)\label{delta_backward}.
\end{align}
The meaning of the different quantities in these equations is as follows:
\begin{itemize}
\item The $\pi N\Delta$ coupling constant is denoted  $f^*=2.13$.

\item 
The $\Delta$-propagator $G_{\alpha\beta}(P)$ is described by  
the Rarita-Schwinger propagator of a spin 3/2 particle
\begin{equation}
 G_{\alpha\beta}(P)= \frac{P_{\alpha\beta}(P)}{P^2-
 M^2_\Delta+i M_\Delta \Gamma_\Delta(P^2)},
\label{delta_propagator}
\end{equation}
where $P_{\alpha\beta}$
is  the projector over
spin-$\frac32$ ,
\begin{align}
P_{\alpha\beta}(P)&=-(\Pbar+M_\Delta)
\left[g_{\alpha\beta}-\frac13\gamma_\alpha\gamma_\beta-
\frac23\frac{P_\alpha P_\beta}{M^2_\Delta}\right.\nonumber\\
&+\left.\frac13\frac{P_\alpha\gamma_\beta-
P_\beta\gamma_\alpha}{M_\Delta}\right]\label{spin32_projector_operator}
\end{align}
and $M_\Delta$ and $\Gamma_\Delta$ 
 stand for the $\Delta(1232)$ resonance
mass and width, respectively.

\item 
In the forward piece, we have introduced the  weak $N\rightarrow\Delta$ 
transition vertex  written as
the sum of vector and axial-vector vertices
\begin{align}
\Gamma^{\beta\mu}(P,Q)&=\Gamma^{\beta\mu}_V(P,Q)
+\Gamma^{\beta\mu}_A(P,Q)
\label{delta_transition_vertex}
\\
\Gamma^{\beta\mu}_V(P,Q)&=\left[\frac{C^V_3}{M}
\left(g^{\beta\mu}\Qbar-Q^\beta\gamma^\mu\right)\right.\nonumber\\
&+\left.\frac{C^V_4}{M^2}\left(g^{\beta\mu}Q\cdot P_\Delta-
Q^\beta P^\mu_\Delta\right)\right. 
\label{vector_delta_transition_vertex}
\\
&+\left. \frac{C^V_5}{M^2}\left(g^{\beta\mu}
Q\cdot P-Q^\beta P^\mu\right)+C^V_6 
g^{\beta\mu}\right]\gamma_5
\nonumber
\\
\Gamma^{\beta\mu}_A(P,Q)&=
\frac{C^A_3}{M}\left(g^{\beta\mu}\Qbar-
Q^\beta\gamma^\mu\right)\nonumber\\
&+ \frac{C^A_4}{M^2}\left(g^{\beta\mu}Q\cdot P_\Delta-
Q^\beta P^\mu_\Delta\right) \nonumber\\
&+ C^A_5 g^{\beta\mu}+\frac{C^A_6}{M^2} 
Q^\beta Q^\mu ,
\label{axial_vector_component}
\end{align}
with $ P_\Delta=P+Q$.

The symbols $C^{V,A}_i$ ($i=3-6$) in the above equations stand for the
$Q^2$-dependent vector and axial-vector form factors.

\item 
In the  backward term we use instead
 the $\Delta\rightarrow N$ transition vertex
given by
\begin{equation}\label{delta_transition_vertex_cc}
\hat{\Gamma}^{\mu\alpha}(P^\prime, Q)=\gamma^0
\left[\Gamma^{\alpha\mu}(P^\prime,-Q)\right]^{\dagger}
\gamma^0.
\end{equation}

\item
The quantities 
$(U^{\rm forw})_{t'_1 t'_2; t_1 t_2}$
and $(U^{\rm back})_{t'_1 t'_2; t_1 t_2}$
 are the matrix elements of the following forward and backward 
isospin operators
\begin{eqnarray}
U^{\rm forw}&=&\left(T_i \left(T^\dagger\right)_{+1}\right)\otimes
\tau_i\\
U^{\rm back}&=&\left(T_{+1}\, T^\dagger_i\right)\otimes
\tau_i,
\end{eqnarray}
where $T_{+1}$ is the spherical 
component of the isovector transition operator 
$\frac32 \rightarrow \frac12$, normalized as
 \begin{equation}\label{Tdaga}
 \left\langle\frac32,t_\Delta\Big|(T^\dagger)_\lambda
 \Big|\frac12,t_N\right\rangle=
C\left(\frac12,1,\frac32\Big|t_N,\lambda,t_\Delta\right)
\end{equation}
for $\lambda=\pm1,0$.

\end{itemize}

\subsection{Isospin structure of MEC}

The isospin dependence of the $\Delta$ current is more complex than
the other operators (seagull, pion-in-flight, and pion-pole). However,
it is possible to expand the $U^{\rm forw,back}$ operators as linear
combinations of the three basic isospin matrices
$\tauvec(1),\tauvec(2)$, and $\Ivec_V$.  This is a consequence of the
following basic property of the isospin transition operators in
cartesian coordinates:
\begin{equation}\label{ttdaga}
T_i \,T^\dagger_j=\frac23\,\delta_{ij}-\frac{i}{3}\,\epsilon_{ijk}\,\tau_k .
\end{equation}

 From this relation it follows that
\begin{align}
\sqrt{3}\,U^{\rm forw}&=
\frac{1}{\sqrt{6}}\left[-2\,\tau_{{}_+}(2)+ 
\left(I_V\right)_{+}\right]\label{U_delta1}\\
\sqrt{3}\,U^{\rm back}&=
\frac{1}{\sqrt{6}}\left[-2\,\tau_{{}_+}(2)- 
\left(I_V\right)_{+}\right]\label{U_deltaprime1}
\end{align}
Analogously, in the  $(1\leftrightarrow2)$ terms of Eqs.~(\ref{delta_forward},\ref{delta_backward})
the isospin operators have to be modified by making the change  
\begin{align}
 \sqrt{3}\,U^{\rm forw}&\quad\underrightarrow{{}^{(1\leftrightarrow2)}}\quad
 \frac{1}{\sqrt{6}}\left[-2\,\tau_{{}_+}(1)- 
\left(I_V\right)_{+}\right]\label{U_delta2}\\
\sqrt{3}\,U^{\rm back}&\quad\underrightarrow{{}^{(1\leftrightarrow2)}}\quad
\frac{1}{\sqrt{6}}\left[-2\,\tau_{{}_+}(1)+ 
\left(I_V\right)_{+}\right],\label{U_deltaprime2}
\end{align}
 where we have made use of the antisymmetry
 property of the isovector operator $\Ivec_V=
 i\left(\tauvec(1)\times\tauvec(2)\right)$ under the 
 interchange $(1\leftrightarrow2)$.

Substituting these relations in Eqs.~(\ref{delta_forward},\ref{delta_backward}),
it is clear that the  $\Delta$-current operator can be written as
the sum of three currents, each one characterized by a specific 
 isospin dependence 
\begin{align}
j^\mu_{\Delta}
&=\tau_{+}(1)\; 
J^\mu_{\Delta 1}(1',2';\,1,2)
+\tau_{+}(2)\; 
J^\mu_{\Delta 2}(1',2';\,1,2)\nonumber\\
&
+\left(I_V\right)_{+}\; 
J^\mu_{\Delta 3}(1',2';\,1,2) ,
\label{deltacur}
\end{align}
where the three functions  $J^\mu_{\Delta i}(1',2';\,1,2)$ depend only
on spins and momenta.

This expression for neutrinos can be applied to antineutrinos by
taking the $(-)$ component of the isospin operators.  In the same way,
for electron scattering one should take the $z$ component of the
isospin operators and keep only the $V$ part of the current.  The
resulting electromagnetic $\Delta$ current is in agreement with
previous expressions~\cite{Amaro:2010iu}.

 From Eqs.~(\ref{seacur},
 \ref{pionic_current1}, 
\ref{polecur},
\ref{deltacur})
we note that the total CC MEC for neutrino scattering 
can be written as
\begin{eqnarray}
 j^\mu_{\rm MEC}
&=&\tau_{+}(1) \,J^\mu_1(1^\prime\,2^\prime;1\,2)+
\tau_{+}(2)\, J^\mu_2(1^\prime\,2^\prime;1\,2)\nonumber\\
&+& \left(I_V\right)_{+}\,J^\mu_3(1^\prime\,2^\prime;1\,2) ,
\label{MEC-isospin}
\end{eqnarray}
where 
\begin{eqnarray}
J_1^{\mu} & = & J_{\Delta 1}^{\mu}\\
J^{\mu}_2 & = & J^{\mu}_{\Delta 2} \\
J^{\mu}_3 & = & 
J^{\mu}_{\rm sea}+J^{\mu}_{\pi}+J^{\mu}_{\rm pole}+J^{\mu}_{\Delta 3}.
\end{eqnarray}

This explicitly shows that the CC MEC operators 
transform as irreducible vectors in
isospin space, implying in particular that 
the final 2p-2h
nuclear states must have $T=1$  for isoscalar nuclei ($T=0$).  
Expression~(\ref{MEC-isospin}) 
will be useful in
obtaining the response functions for the separate charge channels because
the action of the three operators $\tau_+(1)$, $\tau_+(2)$, and
$(I_V)_+$ can be computed directly (see Appendix A).

\section{Electroweak response functions}

In the previous section we  
presented the expressions for the currents in our fully
relativistic model of MEC. These currents were derived from the pion
production amplitudes of \cite{Hernandez:2007qq}.  In this
section we give the explicit expressions for the weak response
functions in the different 2p-2h charge channels.

\subsection{$(\nu_l,l^-)$ responses}

CC neutrino scattering can induce two possible 2p-2h transitions:
 $np \rightarrow pp$ and $nn \rightarrow np$.
In the first case, the $pp$ emission channel, the
diagonal components of the hadronic tensor are of the type
\begin{eqnarray}
W^{\mu\mu}_{pp}
&=&
\frac12 \sumint
\left| \langle pp|
 j^\mu_{MEC}(1'2';12)
\right.
\nonumber\\
&& 
\left.
- j^\mu_{MEC}(2'1';12)|np\rangle 
\right|^2 ,
\label{PP}
\end{eqnarray}
where for brevity we have defined a symbol implying an integration
over momenta and a sum over nucleon spins 
\begin{eqnarray}
\sumint f(1'2';12)
&\equiv&
\frac{V}{(2\pi)^9}
\int
d^3p'_1
d^3h_1
d^3h_2
\frac{M^4}{E_1E_2E'_1E'_2}
\nonumber \\ 
&&
\Theta(p'_1,p'_2,h_1,h_2)
\sum_{s_1s_2s'_1s'_2}
f(1'2';12)
\nonumber \\ 
&&
\delta(E'_1+E'_2-E_1-E_2-\omega),
\end{eqnarray}
where $f(1'2';12)$ is any function depending on the
momenta and spins of the final  2p-2h states.

Note that in the second line of Eq.~(\ref{PP}) we have exchanged the
momenta and spins of the final protons. Here we do not apply the
general Eq.~(\ref{elementary}) which provides the total
elementary hadronic tensor including all the charge channels.

Using the expansion in Eq.~(\ref{MEC-isospin}) we get
\begin{eqnarray}
W^{\mu\mu}_{pp}
&=& 
\frac12 \sumint
\left| \langle pp|
\tau^{(1)}_+ J^\mu_{1}(1'2';12)
+\tau^{(2)}_+ J^\mu_{2}(1'2';12)\right.
\nonumber\\
&&
\left.+(I_V)_+ J^\mu_{3}(1'2';12)\right.
\nonumber\\
&& 
\left.-\tau^{(1)}_+ J^\mu_{1}(2'1';12)
-\tau^{(2)}_+ J^\mu_{2}(2'1';12)
\right.
\nonumber\\
&&
-(I_V)_+ J^\mu_{3}(2'1';12)
|np\rangle\Big|^2 .
\end{eqnarray} 
The isospin matrix elements can be computed 
from Eqs.~(\ref{Ivmas1}--\ref{Ivmas3})
of Appendix A, resulting in
\begin{eqnarray}
W^{\mu\mu}_{pp}
&=& 
2 \sumint
\left| 
J^\mu_{1}(1'2';12)
+J^\mu_{3}(1'2';12)
\right.
\nonumber\\
&&
\left.
-J^\mu_{1}(2'1';12)
-J^\mu_{3}(2'1';12)
\right|^2 .
\end{eqnarray} 
Notice that this is written as the square
 of direct minus exchange matrix elements
of the following ``effective current'' for $pp$-emission with neutrinos
\begin{equation}
J_{pp}^{\mu}= J_1^\mu+J_3^\mu.
\end{equation}
Changing variables $1'\leftrightarrow 2'$ in the final state, it can be
demonstrated that the contribution of the square of the exchange and
direct parts are equal. Thus we obtain
\begin{eqnarray}
W^{\mu\mu}_{pp}
&=& 
4 \sumint
\left\{ 
\left| 
J^\mu_{pp}(1'2';12)
\right|^2
\right.
\nonumber\\
&&
-{\rm Re}\;
J^\mu_{pp}(1'2';12)^*
J^\mu_{pp}(2'1';12)
\Big\} .
\label{wPP}
\end{eqnarray} 
The first term is usually called the ``direct'' contribution, and the
second one is the ``exchange'' contribution, actually being the
interference between the direct and exchange matrix elements.  The
exchange contributions to the 2p-2h cross section have not been included
in the existing models of neutrino scattering~\cite{Nie11,Mar09}, whereas in
this work we include them. In Fig. 2 we show a many-body diagrammatic
representation of the direct and exchange contributions.

\begin{figure}[tph]
\begin{center}
\includegraphics[scale=0.87,bb=200 460 420 720]{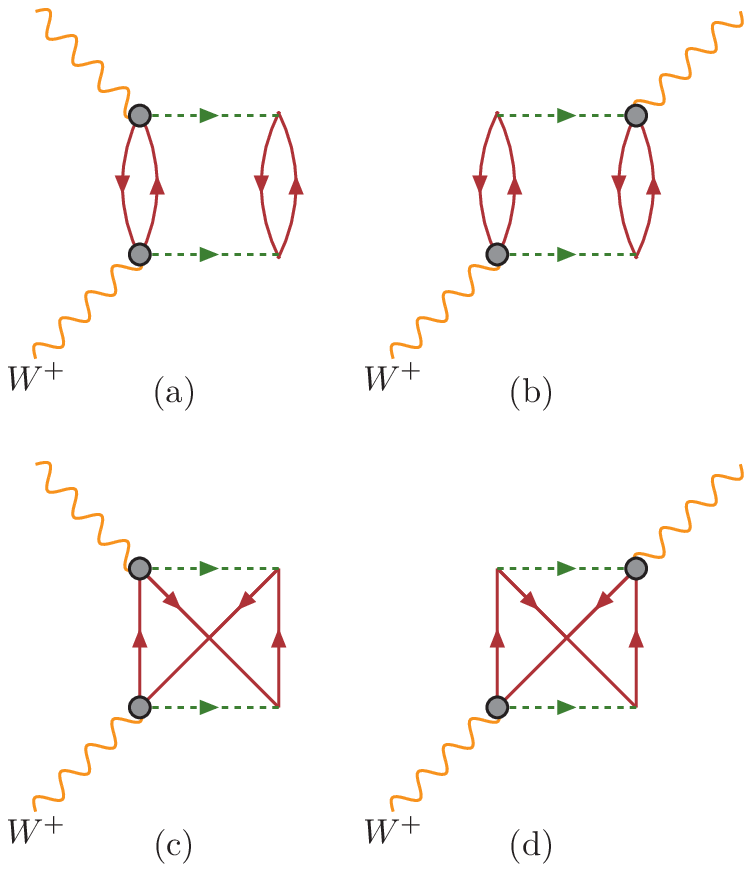}
\caption{Some contributions of 2p-2h to the virtual $W^+$ self-energy,
  or polarization propagator $\Pi^{\mu\nu}$. The response functions
  considered in this work are related to the imaginary part of the
  polarization propagator, ${\rm Im}\; \Pi^{\mu\nu}$.  The circle
  stands for the elementary model for $W^+ N \rightarrow \pi N$ of
  \cite{Hernandez:2007qq} without the nucleon-pole
  diagrams. Diagrams (a,b) represent the direct contribution. Diagrams
  (c,d) are the exchange contributions.  }
\label{selfenergy}
\end{center}
\end{figure}

The $np$ emission case can be obtained in a similar way,
the only difference being that now the exchanged particles 
should be the two initial neutrons. We obtain
\begin{eqnarray}
W^{\mu\mu}_{np}
&=& 
4 \sumint
\left\{ 
\left| 
J^\mu_{np}(1'2';12)
\right|^2
\right.
\nonumber\\
&&
-{\rm Re}\;
J^\mu_{np}(1'2';12)^*
J^\mu_{np}(1'2';21)
\Big\},
\label{wNP}
\end{eqnarray} 
where the effective current for $np$ emission with neutrinos has been defined
\begin{equation}
J^\mu_{np}=J^\mu_2+J^\mu_3.
\end{equation}

\begin{table}
\begin{tabular}{|c|c|c|c|}
\hline
\hfill
final state & $\nu$ & $\overline{\nu}$ & $e$ \\ \hline
$pp$          & $J_1+J_3$ & $\times$ & $J_1+J_2$ \\\hline
$np$          & $J_2+J_3$ & $J_1-J_3$ & $-J_1+J_2$ \\
            &           &           & $2J_3$ \\\hline
$nn$          & $\times$ & $J_2-J_3$ & $-J_1-J_2$ \\ \hline
\end{tabular}
\caption{\label{canales}
Effective currents for two-nucleon emission that appear
in the different charge channels for (anti) neutrinos and electrons (see text).}
\end{table}

The above equations allow one to compute the diagonal hadronic tensor
components appearing in the $LL,CC$ and $T$ responses.  To compute the
$CL$ and $T'$ responses the non-diagonal hadronic tensor components
are necessary. They are computed in a similar way, resulting in
\begin{eqnarray}
W^{\mu\nu}_{pp}
&=& 
4 \sumint
\Big\{ 
J^\mu_{pp}(1'2';12)^*
J^\nu_{pp}(1'2';12)
\nonumber\\
&&
-
J^\mu_{pp}(1'2';12)^*
J^\nu_{pp}(2'1';12)
\Big\}
\label{ndPP}
\\
W^{\mu\nu}_{np}
&=& 
4 \sumint
\Big\{ 
J^\mu_{np}(1'2';12)^*
J^\nu_{np}(1'2';12)
\nonumber\\
&&
-
J^\mu_{np}(1'2';12)^*
J^\nu_{np}(1'2';21)
\Big\}.
\label{ndNP}
\end{eqnarray} 

\subsection{$(\bar{\nu}_l,l^+)$ responses}

In the case of antineutrinos the allowed charge 2p-2h channels are
$np\rightarrow nn$ and $pp\rightarrow np$.  The corresponding formulae
are obtained following the lines of 
the previous section, by taking the matrix
elements of the $(-)$ isospin components of the MEC.  The results are
similar to Eqs.~(\ref{wPP},\ref{wNP},\ref{ndPP},\ref{ndNP}), 
by using the
effective currents given in Table~\ref{canales}.

\subsection{$(e,e')$ responses}

In the case of electron scattering the three charge channels are
all active. We take the matrix elements of the $z$-component of the MEC in
isospin space. In the $pp$ and $nn$ cases the effective currents are
given also in Table~\ref{canales}. For $np$ emission with electrons
two effective currents appear, namely
\begin{eqnarray}
W^{\mu\mu}_{pp}
&=& 
\frac12 \sumint
\left\{ 
\left| 
J^\mu_{pp}(1'2';12)
\right|^2
\right.
\nonumber\\
&&
-{\rm Re}\;
J^\mu_{pp}(1'2';12)^*
J^\mu_{pp}(1'2';21)
\Big\}
\label{emPP}
\\
W^{\mu\mu}_{nn}
&=&W^{\mu\mu}_{pp}
\label{emNN}
\\
W^{\mu\mu}_{np}
&=& 
\sumint
\left\{ 
\left| 
J^\mu_{np1}(1'2';12)
\right|^2
+\left| 
J^\mu_{np2}(1'2';12)
\right|^2
\right.
\nonumber\\
&&
+2{\rm Re}\;
J^\mu_{np1}(1'2';12)^*
J^\mu_{np2}(2'1';12)
\Big\} ,
\label{emNP}
\end{eqnarray} 
where the two effective currents are
\begin{eqnarray}
J^\mu_{np1} &=& -J^\mu_1+J^\mu_2 \\
J^\mu_{np2} &=& 2J^\mu_3.
\end{eqnarray}
These are summarized in fourth column of Table I.

\section{Treatment of the $\Delta$ current}

In this section we provide the details of the treatment of the $\Delta$ current 
in our model and compare with other approaches.

The relativistic $\Delta$ current contribution to the electromagnetic
$R^T$ response was first computed in \cite{Dek94} and \cite{DePace:2003spn}. These authors started with the Peccei
lagrangian for the $\gamma N \Delta$ interaction~\cite{Pec69}. This
introduces a difference with respect to the vector interaction given in
Eq.~(\ref{vector_delta_transition_vertex}). The Peccei vertex only
includes the $O(1/M)$ term that should correspond to the $C_3^V/M$
term of Eq.~(\ref{vector_delta_transition_vertex}).  
There is still
another difference between 
the two approaches because the Peccei vertex
includes a contraction with the tensor
\begin{equation}
\Theta^{\mu\nu}=g^{\mu\nu}-\frac14\gamma^\mu\gamma^\nu .
\end{equation}
This tensor takes into account possible off-shellness effects 
of the virtual $\Delta$~\cite{Pas95}.
In the case the $\Delta$ is on-shell, this is reduced to $g^{\mu\nu}$
because of the properties of Rarita-Schwinger spinors.  

We have verified  that upon multiplying the above tensor by the first
term of Eq.~(\ref{vector_delta_transition_vertex}) the $\Delta$
current of \cite{DePace:2003spn} 
 is reproduced.
This is a consequence of the identity
\begin{equation}
\Theta^{\beta\nu}(g_\nu^\mu\Qbar -Q_\nu\gamma^\mu)
=\frac12 (Q^\mu\gamma^\beta-\gamma^\mu \Qbar\gamma^\beta).
\end{equation}
The resulting current coincides with the one given in \cite{DePace:2003spn} (notice
that there is a relative minus sign with respect to
Eq.~(\ref{delta_propagator}) in the definition of the $\Delta$
propagator in that reference).  We have checked numerically that the
inclusion of $\Theta^{\mu\nu}$ has a negligible effect on the transverse
response at the kinematics relevant for this work, and therefore it
will not be included in the calculations.

Therefore in this work we are using exactly the same operator as in \cite{DePace:2003spn} for the
$\Delta$ vector current, corresponding to the term $C_3^V/M$ in Eq.~(\ref{vector_delta_transition_vertex}). 
We neglect the terms $C_4^V$
and $C_5^V$ ($C_6^V=0$ by conservation of vector current) that are
expected to give much smaller contributions because they are supressed
by $O(P/M)$.  To be consistent, in the axial part we only include the
leading contribution of Eq.~(\ref{axial_vector_component}),
proportional to $C_5^A$ and neglect the other terms.

\begin{figure}[tph]
\begin{center}
\includegraphics[scale=0.87,bb=200 600 420 770]{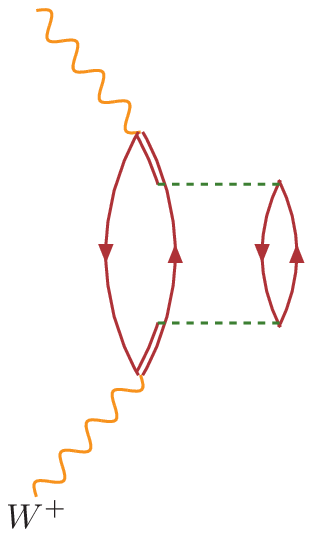}
\caption{Direct term of the MEC 
with excitation of an intermediate $\Delta$
  decaying into a 2p-1h state.  }
\label{delta}
\end{center}
\end{figure}

In what follows we discuss the important
point concerning the ambiguity related to the theoretical separation
between 2p-2h and $\Delta$-peak contributions.  The $\Delta$ peak is the
main contribution to the pion production cross section.  But inside
the nucleus the $\Delta$ can also decay into one nucleon that re-scatters
producing two-nucleon emission without pions. Therefore this decay of
the $\Delta$ should be considered part of the 2p-2h channel. Let us consider
the diagram of Fig.~\ref{delta}. This diagram is implicitly
included in our calculation, being one of the contributions to diagram
(a) of Fig.~\ref{selfenergy}. Thus, it can be considered to contribute 
to the 2p-2h responses and/or to the $\Delta$ peak.  In fact this
diagram contains one self-energy insertion contributing to dressing
the $\Delta$ propagator. As a consequence, there is no unique way of
separating the $\Delta$ emission from the 2p-2h channels because
$\Delta$ emission already includes 2p-2h decays inside the nucleus.

Hence, the MEC contribution given by Eq.~(\ref{hadronic}), while being
purely 2p-2h, also contributes to some extent to the $\Delta$ peak
and, conversely, any calculation of the $\Delta$ peak including a
dressed $\Delta$ propagator would include implicitly some contribution
from the diagram of Fig.~\ref{delta}.  It is a matter of choice in building 
some  specific model whether this contribution is regarded as part of the 
2p-2h or $\Delta$ peak responses. Here we include it in the 2p-2h response.

The previous discussion should make clear that a comparison between
models of MEC that use {\em different prescriptions} for the treatment of the
$\Delta$-pole makes no sense, as long as they contain different admixtures of
$\Delta$ emission. In other words, it is only the total cross section that is meaningful and worthwhile to compare.

Before providing reliable predictions for neutrino scattering, any
model must be validated by confronting it with quasielastic electron
scattering data.  Thus the validation of any prescription for the MEC
contribution requires one to compute the total $(e,e')$ cross section
with a model that includes also both the quasielastic and inelastic
contributions. The validation of our prescription 
for electron scattering has been recently performed in
\cite{Meg16}, where we have shown that the experimental world-data for $^{12}$C can be nicely reproduced within
the super-scaling approach~\cite{SuSAv2} using the MEC of \cite{DePace:2003spn}.
For other models this necessary test has yet to be performed systematically.

\section{Results}

\begin{figure}[ht]
\begin{center}
\includegraphics[scale=0.67,bb=150 115 420 815]{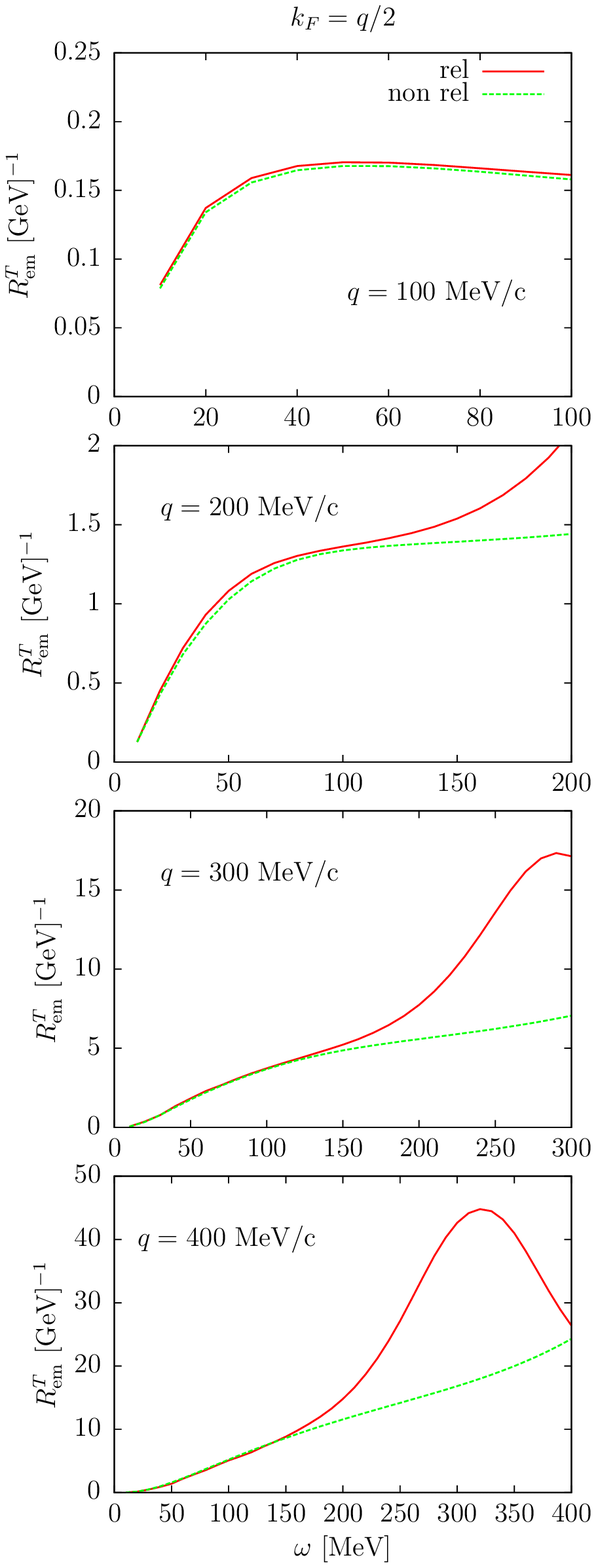}
\caption{Electromagnetic transverse response function for 2p-2h 
for low momentum $q$ and $k_F=q/2$. Here we take $A=56$.
}
\label{low}
\end{center}
\end{figure}

In this section we present results for the five 2p-2h response functions of neutrino
scattering, as a function of $(q,\omega)$. This assumes that the
energy of the incident neutrino is known.  We study the dependence of
the results on several ingredients of the model.

First we validate the relativistic currents for low energy and
momentum transfer by comparison with the electromagnetic transverse response function
 computed in the non-relativistic limit. The relativistic and non-relativistic
responses should coincide in this limit.

We follow the semi-analytical method of \cite{VanOrden:1980tg} (see also
\cite{Alb84,DePace:2003spn}) to compute the non-relativistic
2p-2h transverse response function in electron scattering.  The
comparison with the relativistic calculation also allows one to evaluate
the size of the relativistic corrections. The non-relativistic model is described in Appendix B.

In Fig.~\ref{low} we show the electromagnetic $T$ response for low momentum transfer from $q=100$ to 400 MeV/c and mass number $A=56$. The
value of the Fermi momentum is chosen to be $k_F=q/2$. This is so
because the non-relativistic limit requires that all of the initial and
final momenta go simultaneously to zero, and $k_F$ should also be
reduced accordingly. 
Another reason to reduce $k_F$ in this non-relativistic test
is that for $q < 2 k_F$ Pauli blocking may reduce considerably 
the response function and the comparison cannot be made.

 In the figure we see that for $q=100$ MeV/c the
relativistic and non-relativistic results are the same, and they start
to differ only around $q=200$ MeV/c. The difference is due mainly to the
$\Delta$ propagator, that in one case is considered constant, and in the other case has the
relativistic energy-momentum dependence.  In fact, for these
  low values of $q$, the maximum of the relativistic response
appears around $\omega\sim M_\Delta-M \simeq 300$ MeV, which is the
minimum energy needed to produce the $\Delta$ excitation for a nucleon
at rest.  

\begin{figure}[ht]
\begin{center}
\includegraphics[scale=0.67,bb=150 115 420 815]{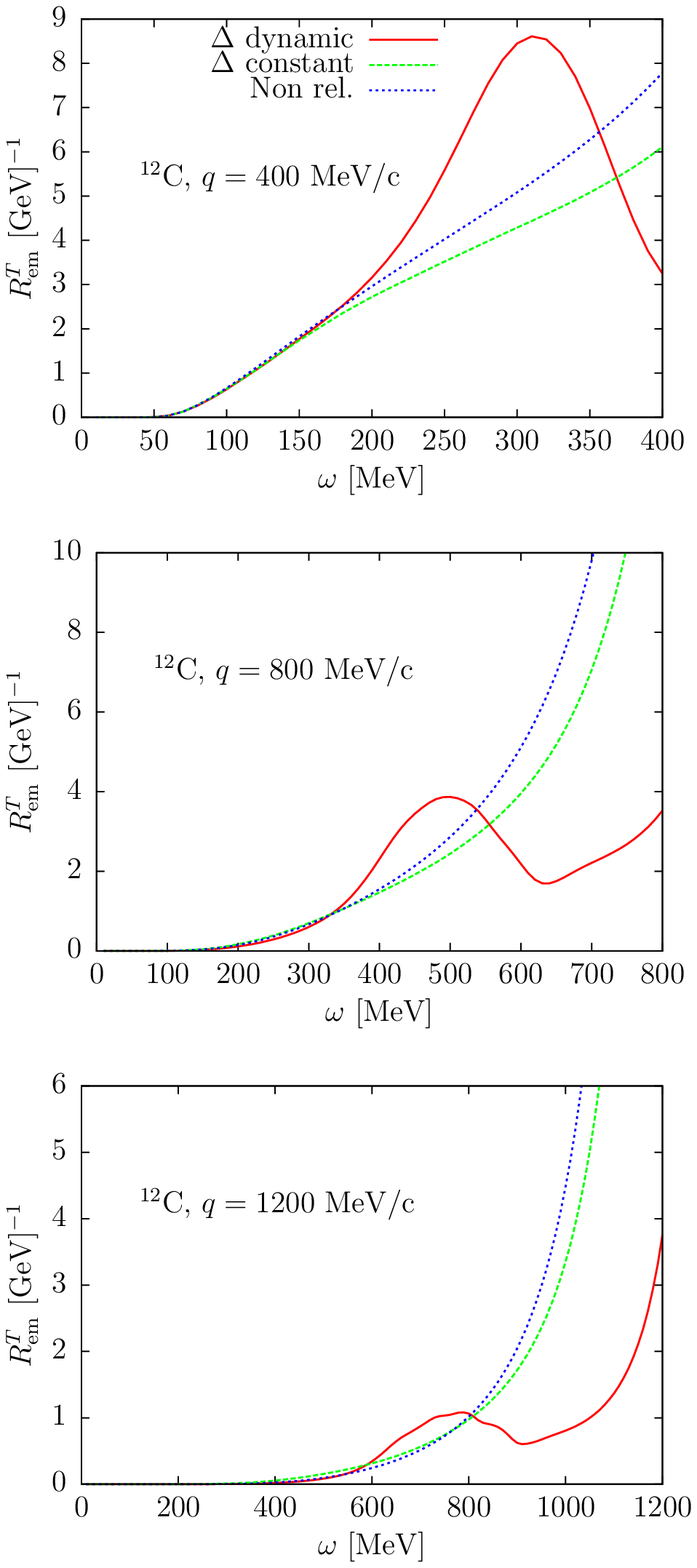}
\caption{Electromagnetic 2p-2h transverse response function of
  $^{12}$C from low to high momentum $q$ and $k_F=228$ MeV/c. We show
  the total relativistic and non-relativistic results, compared to the
  relativistic result with a constant $\Delta$ propagator.  }
\label{static}
\end{center}
\end{figure}

  The
  case $q=400$ MeV/c and above, where $k_F$ takes realistic values, is
  characteristic of what one would expect when one uses a constant
  instead of the dynamical $\Delta$ propagator in the traditional
  non-relativistic calculations. Further insight can be seen in
  Fig.~\ref{static}. The relativistic result with a constant $\Delta$
  propagator is similar to the non-relativistic calculation. In this
  sense the relativistic effects coming from kinematics and spinors
  are smaller than the effects due to the dynamical $\Delta$
  propagator.

\begin{figure}[ht]
\begin{center}
\includegraphics[scale=0.77,bb=200 270 420 780]{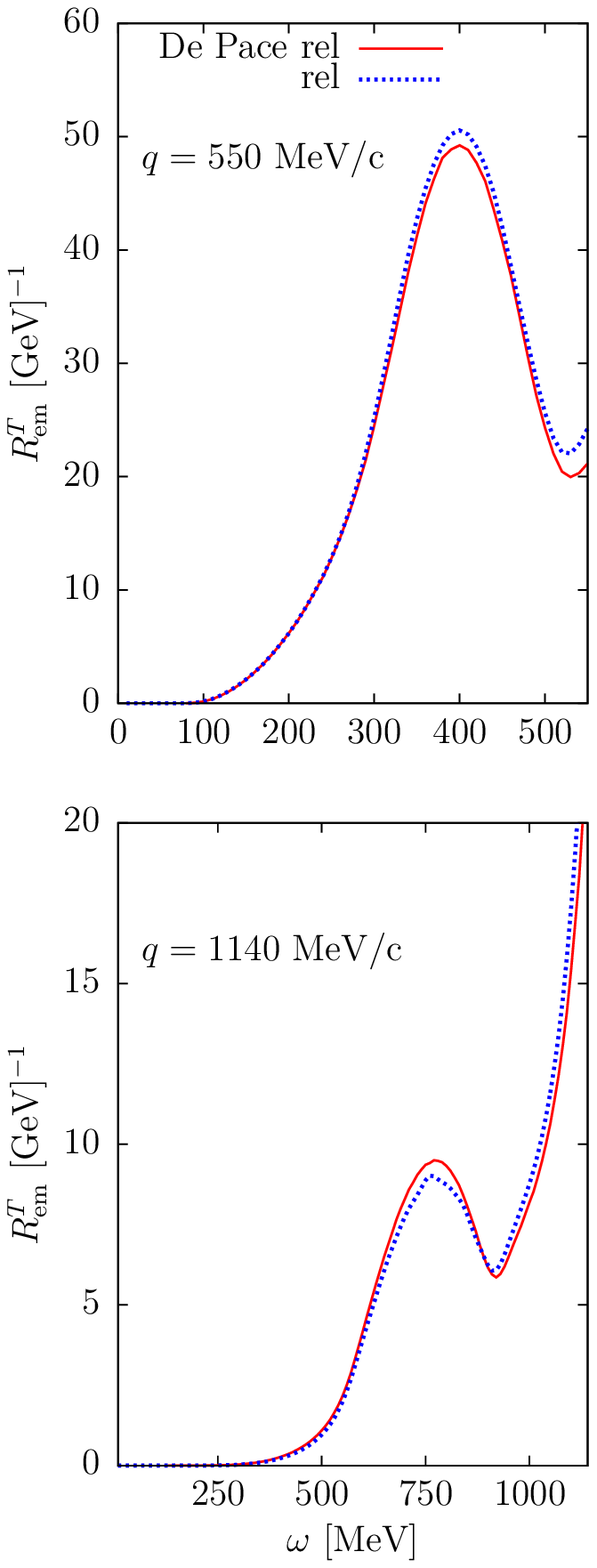}
\caption{Electromagnetic transverse response function for 2p-2h from
  $^{56}$Fe for two values of $q$. Comparison is made with the model of
  \cite{DePace:2003spn}.  }
\label{depace}
\end{center}
\end{figure}

In Fig.~\ref{depace} we compare our results with the calculation of De
Pace {\em et al.}~\cite{DePace:2003spn} for $^{56}$Fe ($k_F=260$ MeV/c), 
including the total MEC
current with direct and exchange contributions.  We use here the same
ingredients as in \cite{DePace:2003spn} 
for the electromagnetic and strong form
factors, and also for the $\Delta$ width.  Only the real part of the
$\Delta$ propagator is included in this calculation, Eq.~(\ref{delta_propagator}).
The two models basically coincide, with only small differences attributed to
the different numerical integration methods used. 
In this sense our model
can be considered as an extension of the model of
\cite{DePace:2003spn} to the charge-changing weak sector.

\begin{figure*}[ht]
\begin{center}
\includegraphics[scale=0.77,bb=30 370 590 780]{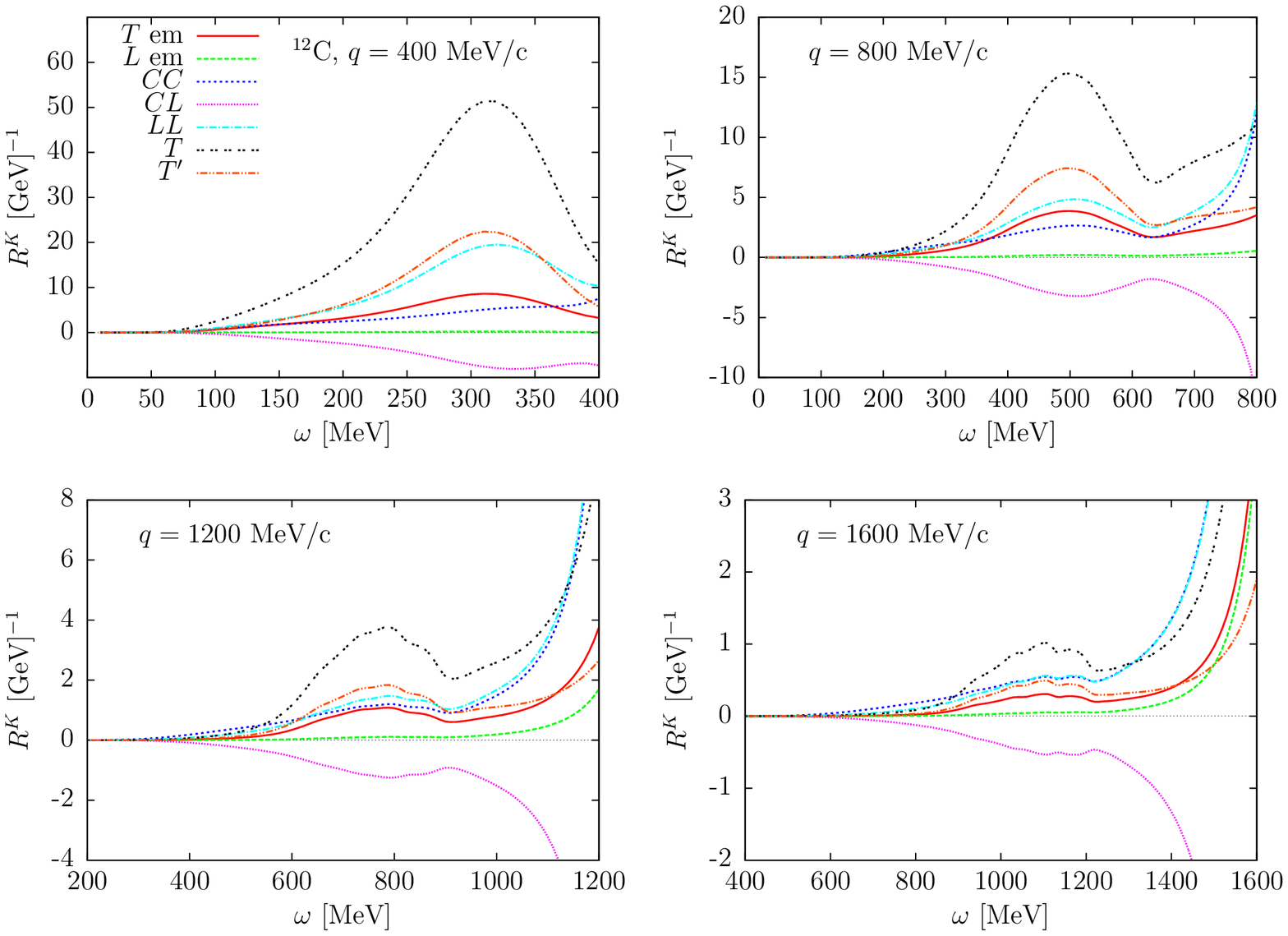}
\caption{Separate 2p-2h response functions of $^{12}$C for four
  values of the momentum transfer. We show the $L$, $T$ electromagnetic
  responses and the five weak responses for charge-changing neutrino scattering. 
To distinguish results for charge-changing neutrino reactions from those for electron scattering we employ the subscript ``em'' for the latter and no subscript for the former.}
\label{responses}
\end{center}
\end{figure*}

In Fig.~\ref{responses} we compare the separate 2p-2h response
functions of $^{12}$C for four values of the momentum transfer.  We
use $k_F=228$ MeV/c and a separation energy $\epsilon=40$ MeV for the
2p-2h state. We show the $L$, $T$ electromagnetic responses and the five
weak responses for CC neutrino scattering.  The transverse response is
dominant because it contains the additive contributions from the $V$ and
$A$ currents. The electromagnetic longitudinal is negligible, but this
is not the case for the neutrino CC response, indicating a large axial
MEC contribution.

Here a few words are 
in order concerning the conventions being
used. 
For electron scattering the transverse em response
has two contributions, one isoscalar and one isovector. For the latter
one typically uses matrix elements of an irreducible tensor operator
in isospin space, for instance, for the one-body current, matrix
elements of $\tau_3=\tau_0$, where $\tau_m$ with $m=0,\pm1$ are the
components of the irreducible tensor. On the other hand, for
charge-changing weak processes it is conventional to use the raising
and lowering operators, which for one-body currents go as
$\tau_{\pm}=\mp \sqrt{2}\tau_{\pm 1}$, giving rise to a factor of 2
between the em isovector transverse response and the CC neutrino
transverse $VV$ response, the latter being twice as large as the former
with these conventions. Of course the em case also has isoscalar
contributions, although these are typically quite small at high
energies where the magnetization current dominates over the convection
current, since the isoscalar to isovector ratio is roughly
$\mu_V^2/\mu_S^2 \simeq 30$.  These arguments are more general and one
finds the same factor for the two-body MEC. In fact, from Eqs.~(\ref{wPP},\ref{wNP}) 
and~(\ref{emPP},\ref{emNN},\ref{emNP}), 
by summing over all the isospin channels, the $VV$ $T$-response is proven to be
twice the em $T$-response.

\begin{figure*}[ht]
\begin{center}
\includegraphics[scale=0.77,bb=30 270 590 780]{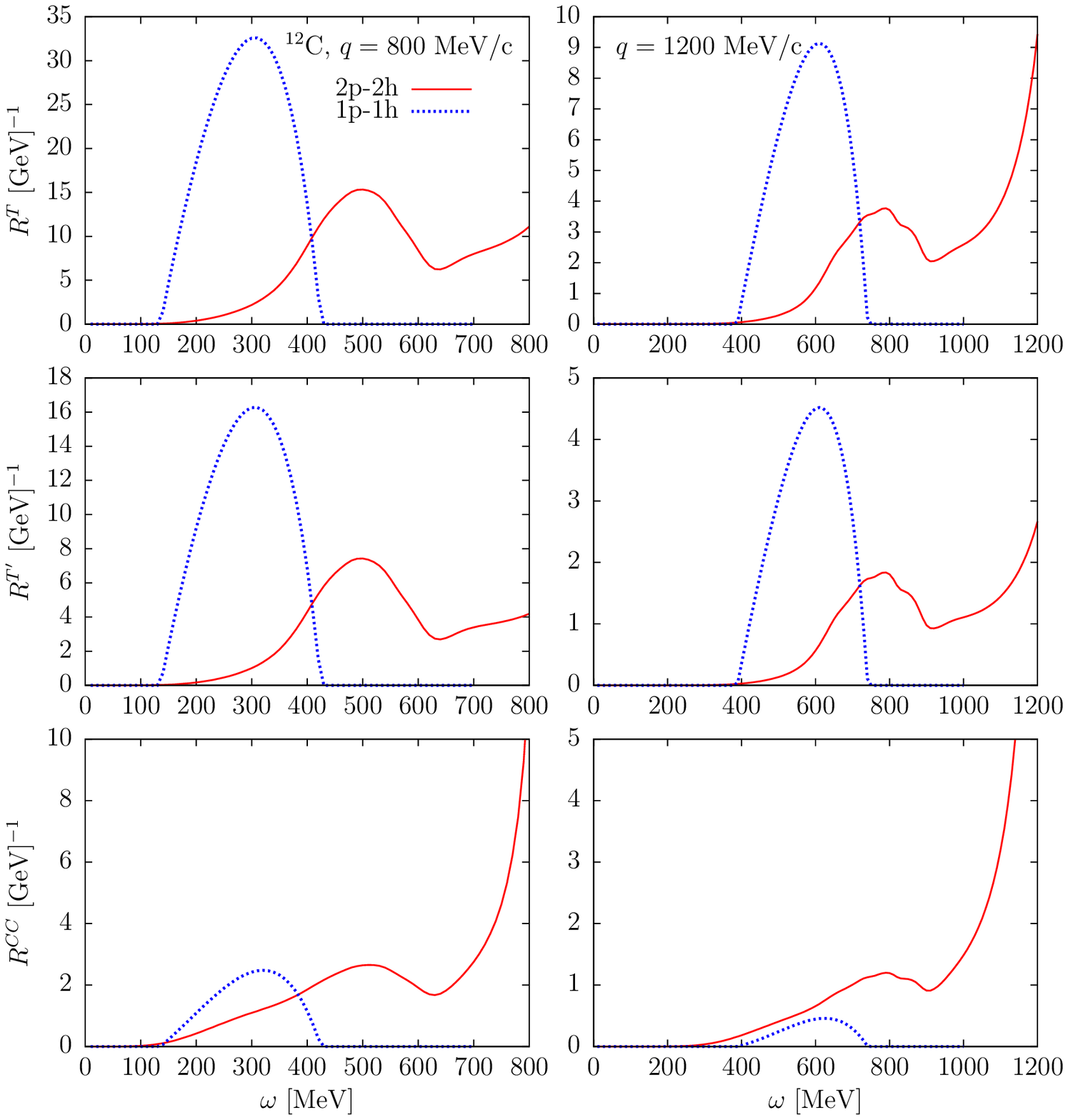}
\caption{Comparison between 1p-1h and  2p-2h response functions
for CC neutrino scattering  off $^{12}$C for two
  values of the momentum transfer. }
\label{neutrino-responses}
\end{center}
\end{figure*}

\begin{figure*}[ht]
\begin{center}
\includegraphics[scale=0.77,bb=30 450 590 780]{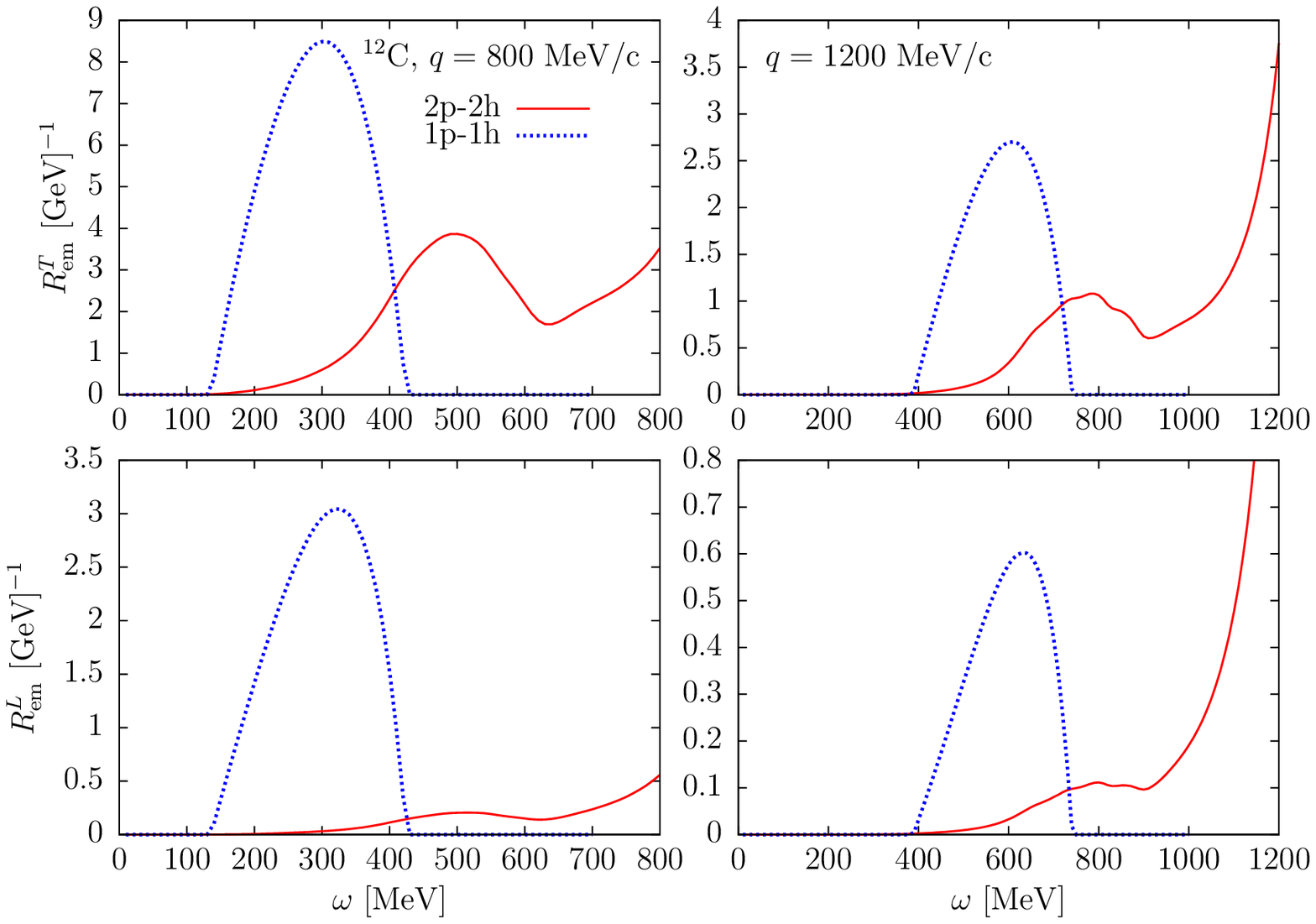}
\caption{Comparison between 1p-1h and  2p-2h response functions
for electron scattering  off $^{12}$C for two
  values of the momentum transfer. 
 }
\label{electron-responses}
\end{center}
\end{figure*}

In Fig.~\ref{neutrino-responses} we compare the 1p-1h and 2p-2h
neutrino responses for $q=800$ and 1200 MeV/c. The 1p-1h responses are
computed in the RFG and only contain the one-body (OB) current. For
these values of $q$ there are large MEC effects. The 2p-2h strength at
the maximum of the $\Delta$ peak is around $50\%$ of the 1p-1h
response. The MEC effects are similar in the $T$ and $T'$ responses.
The MEC effects in the $CC$ response are relatively much larger than in
the transverse ones. This indicates again a large longitudinal
contribution of the axial MEC.

It should be kept in mind, however, that each response function
appears in the cross section multiplied by a kinematical factor (see
Eq.~(\ref{Spm})) which alters the balance shown in Fig.~\ref{responses}. For example,
the contributions of the three responses $R^{CC}$, $R^{CL}$ and $R^{LL}$
largely cancel each other, yielding a net charge/longitudinal cross
section that is generally smaller than the transverse ones. This balance of the 
different response functions of course depends on the kinematics.

For comparison, in Fig.~\ref{electron-responses} we show the
electromagnetic response functions. Here the MEC effects in the
transverse case are similar to those shown for neutrinos, while the
longitudinal MEC contribution is very small except as one approaches the lightcone $(Q^2=0)$ where they become large. One should remember, however, that the kinematic factor $v_L$ that multiplies the longitudinal response goes to zero in the real-photon limit.

\begin{figure}[ht]
\begin{center}
\includegraphics[scale=0.77,bb=170 160 440 780]{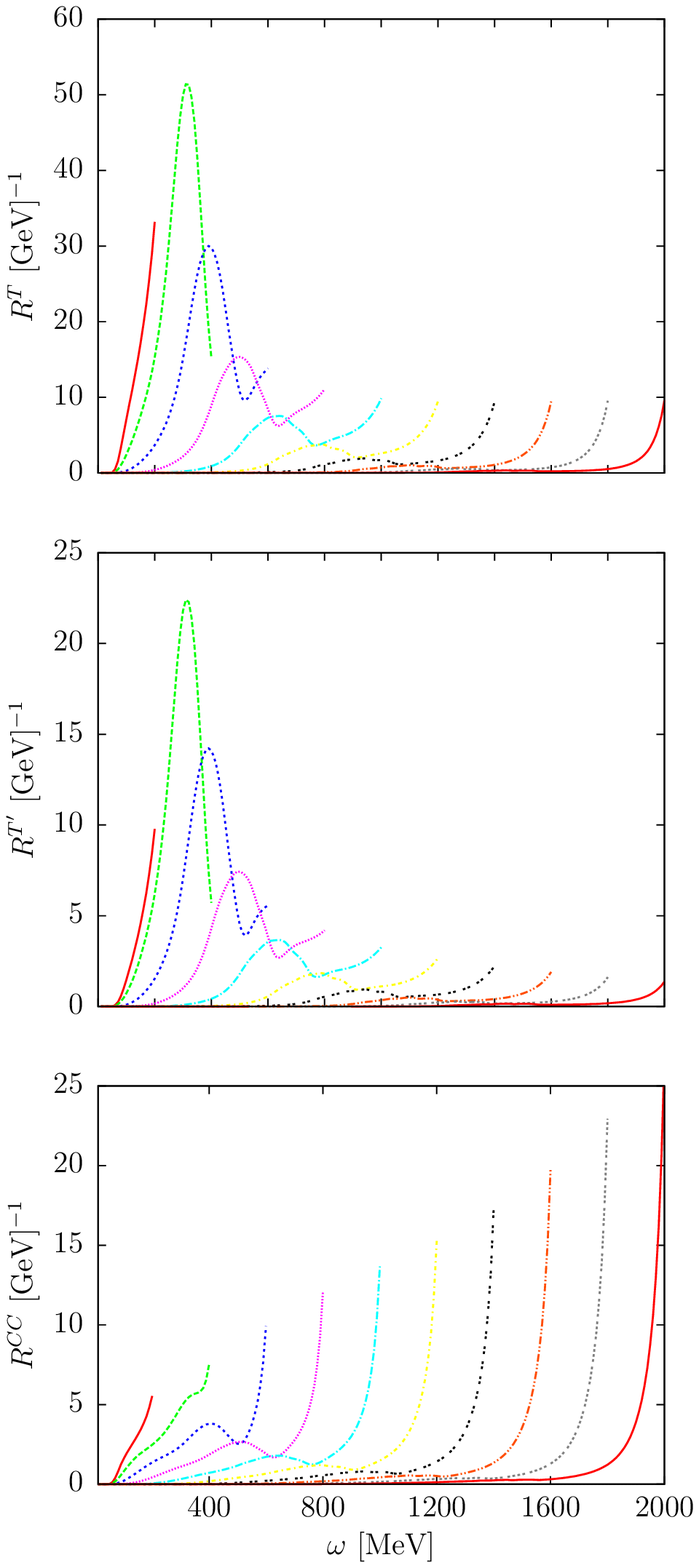}
\caption{Evolution of the weak 2p-2h response functions from low to
  high values of $q$. Only the $T$, $T'$ and $CC$ responses are shown
  from left to right, for $q=200$, 400, 600, 800, 1000, 1200, 1400,
  1600, 1800 and 2000 MeV/c.  }
\label{evolution}
\end{center}
\end{figure}

In Fig.~\ref{evolution} we show the behavior of three of the weak
response functions from low to high $q$. The MEC peak moves from left
to right according approximately to the $\Delta$ position $\omega \sim
\sqrt{M_\Delta^2+q^2}-M$, and its strength decreases due to the form factors, 
since $Q^2$ at the peak position increases with $q$. 
While one might be tempted to conclude that the MEC contributions become negligible at high $q$, it should be remembered that the QE response also decreases as
$q$ increases.
In fact, a better representation of the relative importance of these two contributions can be obtained by forming the so-called reduced response used in scaling analyses, {\it i.e.}, by dividing the responses by the single-nucleon expression that makes the QE response scale (see \cite{Donnelly:1998xg,Meg15}). 
In this representation the QE contribution plotted versus the corresponding scaling variable becomes universal --- a single curve is obtained.
Doing the same for the MEC contribution yields a better understanding of the relative importance of the two contributions.
In fact, in going from low $q$ to 2000 MeV/c the MEC reduced response falls only by about a factor of two (see also \cite{Meg15}).
The $T$ and $T'$
    responses are similar in shape and the size of $R^T$ is around
    twice that of $R^{T'}$.

\begin{figure}[ht]
\begin{center}
\includegraphics[scale=0.77,bb=170 160 440 780]{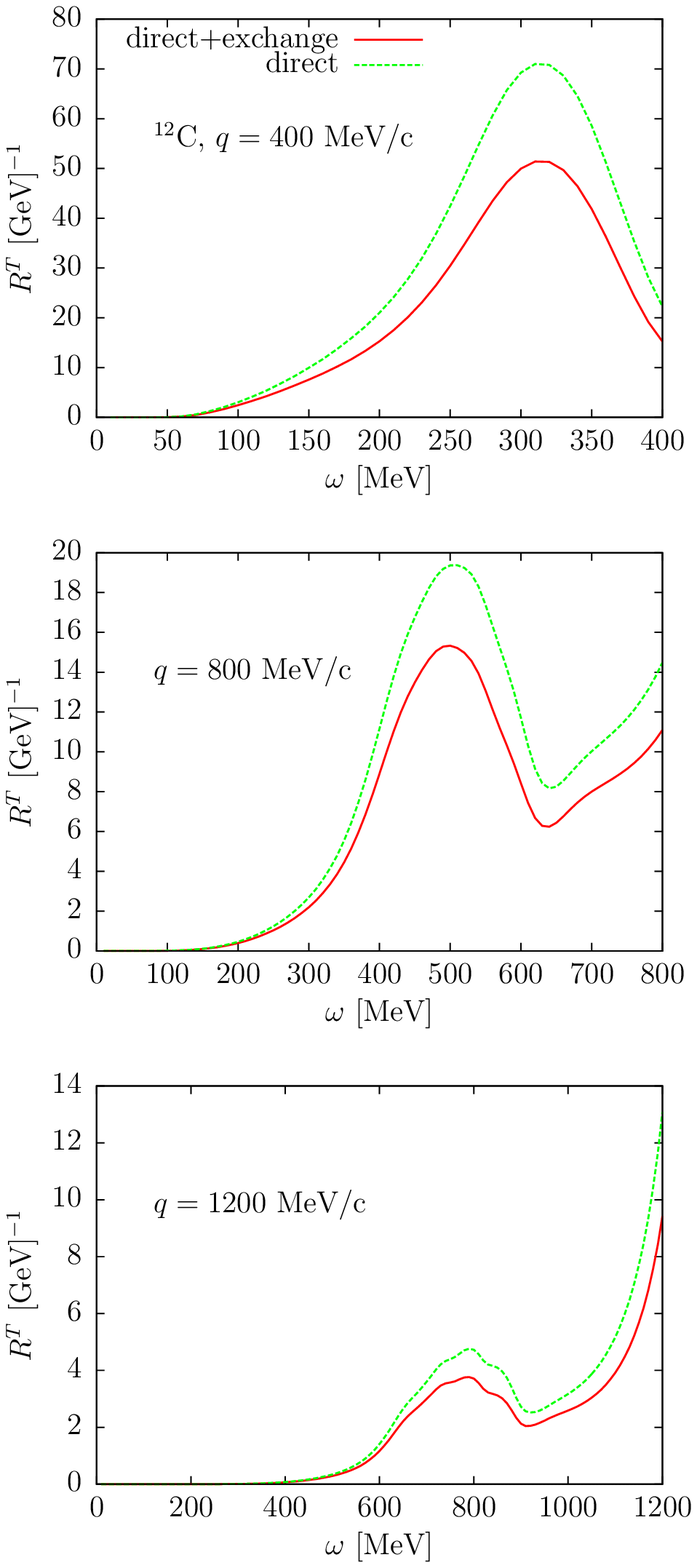}
\caption{The $T$ weak response function computed including only the direct
  contribution compared to the direct+exchange contribution.  }
\label{exchange}
\end{center}
\end{figure}

In Fig.~\ref{exchange} we show the effect of neglecting the exchange
contribution (see diagrams (c), (d) of Fig. 2). This amounts to about
a 25\% increase. This is in agreement with \cite{DePace:2003spn} 
and also with previous studies of the exchange
pieces in the $\Delta$ self-energy~\cite{Ose87}. We conclude, as in \cite{DePace:2003spn}, that the exchange contribution is not negligible.

\section{Conclusions and perspectives}

In this work we have presented a fully relativistic model of electroweak
meson-exchange currents for inclusive CC neutrino scattering,
which is an extension of the relativistic electron scattering MEC model of \cite{DePace:2003spn}. 
The currents have been derived from the pion
production amplitudes of \cite{Hernandez:2007qq}.  

We have given expressions for the 2p-2h response functions in the
relativistic Fermi gas model for the different charge channels. We
have presented results for the response functions from low to large
$q$ values $q=200 \ldots 2000$ MeV/c. Our calculation has no
approximations and we compute the full 7D integrals including the
exchange contributions. We have studied the dependence of the results
on different ingredients of the model, and made comparisons with the 1p-1h
channel.

We have found large effects due to the dynamical  character of the $\Delta$ propagator. Moreover, we have shown that,
although the transverse responses dominate, the
longitudinal ones are not negligible -- as they are in electron
scattering -- due to the presence of a large longitudinal axial
component. The 2p-2h states are found to be important compared with the 1p-1h ones
for all kinematics and all response functions. In particular
the MEC effect is very large in the longitudinal responses, 
although the impact of this on the cross section depends on the kinematics.

We have discussed some important issues concerning the relativistic
two-body $\Delta$ current and possible double-counting problems. In
this work we have kept only the real part of the $\Delta$ propagator
to avoid a large contamination from the $\Delta$ pion emission
peak. The inherent uncertainty related to this approach has been
discussed and remains to be quantified.

Finally we have studied the effects of neglecting the exchange
contribution of the MEC for neutrino scattering 
and we have found that they amount to about +25\%, in
agreement with what was found for electron scattering by other authors~\cite{DePace:2003spn}.

In future work we will provide predictions for the neutrino cross
sections.  Including the neutrino flux in the present model for the
calculation of the 2p-2h neutrino cross section would imply performing an
8-dimensional integration, increasing considerably the computational
time. To compute flux-integrated cross sections it is more practical
to resort to a parametrization of the response functions, 
allowing one to perform
the additional integration over the neutrino energy distribution~\cite{Meg15,Meg16}: such parametrizations will be provided in the near future.
Alternatively, one may invoke some type of approximation as noted below.

In particular it is interesting to study the validity of the frozen
nucleon approximation introduced in \cite{Ruiz14,Ruiz14b,Simo2015}
for the phase space, including the MEC operators. This approximation
reduces the integration to one dimension, and the code can be swiftly 
implemented in Monte Carlo event generators.

\section*{Acknowledgments}
This work was supported by Spanish Direccion General de
Investigacion Cientifica y Tecnica and FEDER funds (grants
No. FIS2014-59386-P and No. FIS2014-53448-C2-1), by the Agencia de
Innovacion y Desarrollo de Andalucia (grants No. FQM225, FQM160), by INFN under
project MANYBODY, and part (TWD) by U.S. Department of Energy under
cooperative agreement DE-FC02-94ER40818. IRS acknowledges support
from a Juan de la Cierva fellowship from Spanish MINECO. 
J.E.A. and I.R.S. thank E. Hernandez and J.M. Nieves for useful
discussion on the pion production model.


\appendix

\section{Matrix elements of isospin operators}\label{appendix_A}

In this work we follow the convention in which the proton isospin state 
$\left|p\right\rangle$ corresponds to isospin projection
$t_z=+\frac12$, and the neutron one $\left|n\right\rangle$ 
corresponds to $t_z=-\frac12$.

The operators $\left(I_V\right)_{\pm}$, appearing in the corresponding
formulae for neutrinos and antineutrinos, couple to the $W^\pm$,
respectively.
The $+$ component reads
\begin{equation}
(I_V)_+ = \tau_+\otimes\tau_z -\tau_z\otimes\tau_+ ,
\end{equation}
where $\tau_+=\tau_x+i\tau_y$.
  From this expression, the operation on a nucleon pair gives
\begin{eqnarray}
\left(I_V\right)_{+}\left|np\right\rangle&=&
2\left|pp\right\rangle
\label{Ivmas1}\\ 
\left(I_V\right)_{+}
\left|pn\right\rangle&=&
-2\left|pp\right\rangle
\label{Ivmas2}\\ 
\left(I_V\right)_{+}
\left|nn\right\rangle&=& 2\left|np\right\rangle
-2\left|pn\right\rangle
\label{Ivmas3}\\
\left(I_V\right)_{+}
\left|pp\right\rangle&=& 0 .
\label{Ivmas4}
\end{eqnarray}
Interchanging protons and neutrons the action of $(I_V)_-$ is readily
obtained as
\begin{eqnarray}
\left(I_V\right)_{-}\left|pn\right\rangle&=&
2\left|nn\right\rangle \label{Ivmenos1}\\ \left(I_V\right)_{-}
\left|np\right\rangle&=&
-2\left|nn\right\rangle\label{Ivmenos2}\\ \left(I_V\right)_{-}
\left|pp\right\rangle&=& 2\left|pn\right\rangle\label{Ivmenos3}
-2\left|np\right\rangle\\ \left(I_V\right)_{-}
\left|nn\right\rangle&=& 0 .
\end{eqnarray}
This is a consequence of the underlying isospin symmetry of the weak
interaction.

Note that in the electromagnetic MEC,
the isospin operator $\left(I_V\right)_z$ 
is always associated with the interchange
of a charged pion ($\pi^\pm$) between two nucleons. 
This is due to the fact that the $\gamma\pi$NN vertex
comes from the pseudo-vector $\pi$NN coupling
with the prescription of electromagnetic minimal coupling. This 
automatically ensures charge conservation and only couples
the electromagnetic potential to charged particles.

Therefore, the action of this operator will always imply a charge
interchange between the two nucleon species in the initial 
state. Its effect on the states where both nucleons have  
well-defined 3rd-components
of the individual isospins (uncoupled basis)
$\left| N_1 N_2\right\rangle$ is
\begin{eqnarray}
\left(I_V\right)_z \left|pn\right\rangle&=& 2\left|np\right\rangle\\
\left(I_V\right)_z \left|np\right\rangle&=& -2\left|pn\right\rangle
\end{eqnarray}
and zero when acting on the other two states, $\left|pp\right\rangle$
and $\left|nn\right\rangle$.

\section{Non-relativistic approach}\label{subsec:norel}

The hadronic tensor for the elementary 2p-2h transition,
Eq.~(\ref{elementary}), contains the direct and exchange matrix
elements of the two-body current operator. If one neglects the
interference between the direct and exchange terms,
in the non-relativistic case, 
$r^{\mu\nu}$ is a function of $k_1,k_2$
only, where $\nk_i= \np'_i-\nh_i$,
or equivalently, a function 
of the dimensionless variables 
\begin{equation}
x= k_1/k_F, \kern 1cm
y= k_2/k_F.
\end{equation}
Then the following~\cite{VanOrden:1980tg} change of variables 
\begin{eqnarray}
\nl_1= \frac{\np'_1-\nh_1}{k_F}
&&
\nl_2= \frac{\np'_2-\nh_2}{k_F}
\\
\nx_1= \frac{\np'_1+\nh_1}{2 k_F}
&&
\nx_2= \frac{\np'_2+\nh_2}{2 k_F}
\end{eqnarray}
allows one to compute analytically the integral over $\nx_1,\nx_2$
given by the function
\begin{eqnarray}
A(l_1,l_2,\nu)
&=&
\frac{l_1^3l_2^3}{(2\pi)^2}
\int d^3x_1d^3x_2
\delta(\nu-\nl_1\cdot\nx_1-\nl_2\cdot\nx_2)
\nonumber\\
&&
\theta\left( 1-\left|\nx_1-\frac{\nl_1}{2}\right| \right)
\theta\left( 1-\left|\nx_2-\frac{\nl_2}{2}\right| \right)
\nonumber\\
&&
\theta\left( \left|\nx_1+\frac{\nl_1}{2}\right| -1 \right)
\theta\left( \left|\nx_2+\frac{\nl_2}{2}\right| -1 \right),
\nonumber\\
&&
\end{eqnarray}
where $\nu = M\omega/k_F^2$. 
This function has been computed analytically in \cite{VanOrden:1980tg}
and more recently  in \cite{Ruiz14},
in relation to a typo in one of the terms in the original reference.
The 2p-2h transverse response is
\begin{eqnarray}
R^T_{2p-2h}(q,\omega) &=&
\frac{V}{(2\pi)^6}\frac{k_F^7\, M}{q_F}
\int_0^{x_{\rm max}}
\frac{dx}{x^2}
\int_{|q_F-x|}^{q_F+x}
\frac{dy}{y^2}
\nonumber\\
&&\times
 A(x,y,\nu)
r^{T}(x,y) ,
\nonumber\\
\label{fasico2D}
\end{eqnarray}
where the upper limit is 
$x_{\rm max} =  1 + \sqrt{2(1+\nu)}$.
We have also defined the following dimensionless variable
\begin{eqnarray}
\nq_F &=& \frac{\nq}{k_F} .
\end{eqnarray}
The two-dimensional integral above has to be performed
numerically. This integral has been studied in \cite{Ruiz14}.

The elementary 2p-2h response $r^T(x,y)$ was computed in
\cite{VanOrden:1980tg} by performing the spin-isospin traces of the
non-relativistic MEC. In this work we have derived again the analytical
expressions for the traces and have detected some typos in
Eqs. (2.24) and (2.25) of that reference. For completeness we write
here the correct expressions. 
Note that, despite these typographical errors, the numerical results of 
\cite{VanOrden:1980tg} appear to be correct.

We write the total response as the sum of seagull, pion-in-flight and  
pure $\Delta$ responses plus their interferences
\begin{equation}
r^T= r^T_{\rm sea}+r^T_{\rm \pi}+r^T_{\rm \Delta}+r^T_{\rm
  sea,\pi}+r^T_{\rm sea,\Delta}+r^T_{\rm \pi,\Delta} .
\end{equation}
The different contributions are
\begin{eqnarray}
r^T_{\rm sea}(x,y)
&=&
\left(2\frac{f_{\pi NN}^2}{m_\pi^2}F_1^V\right)^2
\frac{8}{k_F^2}
\left[
\frac{x^2}{(x^2+m_F^2)^2}
\right. \nonumber
\label{rtxy}\\ 
&&
\left. 
+\frac{y^2}{ (y^2+m_F^2)^2 }
+\frac{x_T^2}{ (x^2+m_F^2)(y^2+m_F^2)  }
\right] \,,
\nonumber \\
&&
\end{eqnarray}
where $m_F \equiv m_{\pi}/k_F$ and
\begin{equation}
x_T^2= x^2 -\left( \frac{q_F^2+x^2-y^2}{2q_F} \right)^2 \,.
\end{equation}
This non-relativistic 
result coincides with~\cite{VanOrden:1980tg}
 for the seagull current 
except for the last plus sign in 
the last term of Eq.~(\ref{rtxy}) that in the cited reference is a 
multiplication sign (see Eq.~(2.24) of~\cite{VanOrden:1980tg}).

The pion-in-flight response and its interference with the seagull current are
\begin{eqnarray}
r^T_{\rm \pi}(x,y)
&=&
\left(2\frac{f_{\pi NN}^2}{m_\pi^2}F_1^V\right)^2
\frac{16}{k_F^2}
\frac{x^2y^2x_T^2}{ (x^2+m_F^2)^2(y^2+m_F^2)^2  }
\nonumber\\
\\
r^T_{\rm sea,\pi}(x,y)
&=&
-\left(2\frac{f_{\pi NN}^2}{m_\pi^2}F_1^V\right)^2
\frac{16}{k_F^2}
\frac{x_T^2}{ (x^2+m_F^2)(y^2+m_F^2)  }
\nonumber\\
&\times&
\left[
\frac{x^2}{(x^2+m_F^2)}
+\frac{y^2}{ (y^2+m_F^2) }
\right]. 
\end{eqnarray}
This result coincides with~\cite{VanOrden:1980tg}, except for the 
$x^2$-term in the interference, which was missing in 
 Eq.~(2.24) of~\cite{VanOrden:1980tg}.

In the case of the $\Delta$ current various schemes are typically adopted in going to the non-relativistic limit and thus
in deriving the non-relativistic limit of the
$\Delta$ current.
We take the approach described in
\cite{Ama03}, where the non-relativistic reduction of the 
$\Delta$ current reads
\begin{eqnarray}
{\bf J_\Delta} 
&=&
\frac{i}{6}\frac{C_3^V f^* f_{\pi NN}}{M m_\pi^2}
\frac{\nk_2\cdot\nsigma^{(2)}}{m_\pi^2+k_2^2}
\left[ B\tau_z^{(2)}\nk_2 
\right.
\nonumber\\
&&
\left.
-\frac{A}{2}
[\ntau^{(1)}\times\ntau^{(2)}]_z 
\nsigma^{(1)}\times\nk_2\right]\times\nq 
\nonumber\\
&& + (1\leftrightarrow 2),
\end{eqnarray} 
where $A=\frac{8}{3(M_\Delta-M)}$ and $B=2A$.
The pure $\Delta$ response can be written as
\begin{eqnarray}
r^T_\Delta
&=&
\frac{4}{k_F^2}
\left(2\frac{f_{\pi NN}^2}{m_\pi^2}\right)^2
\left[
q_F^2a^2k_F^4
\left( 
\frac{x^2(x^2+x_L^2)}{ (x^2+m_F^2)^2 }
\right.
\right.
\nonumber\\
&&
\left.
+\frac{y^2(y^2+y_L^2)}{ (y^2+m_F^2)^2 }
+\frac{2q_F^2x_T^2}{ (x^2+m_F^2)(y^2+m_F^2) }
\right)
\nonumber\\
&+&
\left.
2q_F^2b^2k_F^4x_T^2
\left( 
\frac{x^2}{ (x^2+m_F^2)^2 }
+\frac{y^2}{ (y^2+m_F^2)^2 }
\right)
\right] ,
\nonumber\\
\label{delta-delta}
\end{eqnarray}
where we have defined 
\begin{equation}
a= \frac{1}{2}\frac{C_3^V}{M}\frac{f^*}{6f_{\pi NN}}A
\end{equation}
and $b=2a$. The $a,b$ factors notation is similar to that used in
\cite{VanOrden:1980tg}, while the $A, B$ factors are used in 
\cite{DePace:2003spn}. Note that there are several
definitions for $A$ and $B$ in the literature which arise from
different approximations in deriving the non-relativistic limit of the
$\Delta$ current, already discussed in \cite{DePace:2003spn}. 
The expression for $a$, $b$, written in terms of
$A$ and $B$ correspond to Eq.~(2.25) of \cite{VanOrden:1980tg},
where there is a typo in the denominator. What should appear is $M$ instead
of $M_\Delta$.

Now we have also defined
\begin{eqnarray}
x_L & = &  \frac{q_F^2+x^2-y^2}{2q_F}
\\
y_L & = &  \frac{q_F^2+y^2-x^2}{2q_F} .
\end{eqnarray}
Note the opposite definition in the sign of $x_L$ with respect \cite{VanOrden:1980tg}.

In Eq.~(2.24) of \cite{VanOrden:1980tg} $x_T^2$ was globally
factorized from the first two lines of Eq.~(\ref{delta-delta}), while
$k_F^2$ was written in the third line instead of $k_F^4$.
Finally, the interference between $\Delta$ and seagull and 
pion-in-flight currents are given by 
\begin{eqnarray}
r^T_{\rm sea,\Delta}
&=&
-\frac{4}{k_F^2}
\left(2\frac{f_{\pi NN}^2}{m_\pi^2}\right)^2
F_1^V
4a q_Fk_F^2
\left[
\frac{x^2x_L}{ (x^2+m_F^2)^2 }
\right.
\nonumber\\
&&
\left.
+\frac{y^2y_L}{ (y^2+m_F^2)^2 }
+\frac{q_Fx_T^2}{ (x^2+m_F^2)(y^2+m_F^2)}
\right]
\\
r^T_{\rm \pi,\Delta}
&=&
\frac{4}{k_F^2}
\left(2\frac{f_{\pi NN}^2}{m_\pi^2}\right)^2
F_1^V
4a k_F^2
\frac{q^2_Fx_T^2}{ (x^2+m_F^2)(y^2+m_F^2)}
\nonumber\\
&&
\left[
\frac{x^2}{ (x^2+m_F^2) }
+\frac{y^2}{ (y^2+m_F^2) }
\right] .
\end{eqnarray}


\end{document}